\documentclass[%
 aip,
 pof,%
 amsmath,amssymb,
 reprint,%
]{revtex4-1}
\usepackage {CJK}
\usepackage{graphicx}
\usepackage{dcolumn}
\usepackage{bm}
\usepackage[mathlines]{lineno}
\usepackage{color}
\usepackage{gensymb}
\usepackage{hyperref}

\begin{document}
\begin{CJK*}{GB}{gbsn}

\title{Crown rupture during droplet impact on a dry smooth surface at increased pressure}
\author{Zhigang Xu (徐志刚)}
\affiliation{State Key Laboratory of Engines, Tianjin University, Tianjin, 300072, China.}%
\author{Longlong Wang (王龙龙)}
\affiliation{State Key Laboratory of Engines, Tianjin University, Tianjin, 300072, China.}%
\author{Tianyou Wang (王天友)}
\affiliation{State Key Laboratory of Engines, Tianjin University, Tianjin, 300072, China.}%
\author{Zhizhao Che (车志钊)}
\email{chezhizhao@tju.edu.cn}
 \affiliation{State Key Laboratory of Engines, Tianjin University, Tianjin, 300072, China.}

\date{\today}

\begin{abstract}
The impact of droplets at increased environmental pressure is important in many industrial applications. Previous studies mainly considered the impact process at standard or reduced environmental pressure, and the effect of high environmental pressure is unclear. In this study, we experimentally investigate the impact of ethanol droplets on dry smooth surfaces at increased environmental pressure. The effects of the environmental pressure on the splashing and rupture of the crown during the impact process are analyzed. The results show that surrounding gas with high environmental pressure can lead to the splashing of the crown in a `thread rupture' mode and the sizes of the secondary droplets from the rim of the liquid crown increase with the environmental pressure. The threshold for the transition from spreading to splashing during the impact process is obtained based on the theory of aerodynamics analysis of the lamella. At increased environmental pressure, the threshold speed of the impact decreases with increasing the environmental pressure because the wedge of the lamella is prevented from moving forward and is driven to detach from the substrate by the air ahead, which has a higher density due to the higher environmental pressure.
\end{abstract}


\maketitle
\end{CJK*}

\section{Introduction}\label{sec:sec01}
Droplet impact is of great significance for a wide variety of industrial processes, such as spray combustion \cite{Wang2004FilmBreakup}, spray coating \cite{Bergeron2000NatureImpactPolymer}, spray cooling \cite{Yarin2006RecoilBouncing}, and pesticide deposition \cite{Wirth1991AgricultureSpray}. Most studies on the impact of droplets were conducted at the standard environmental pressure until Xu et al.\ \cite{Xu2005Splashing}, who first revealed the importance of the environmental pressure on the impact dynamics by performing the impact experiments below the atmospheric pressure and found that splashing could be suppressed by reducing the environmental pressure. Since that, several studies of droplet impact \cite{Xu2007SurroundingGas, Latka2012, Li2017DoubleContact, Latka2017ThresholdPressures} have been conducted by decreasing the environmental pressure. For example, Xu et al.\ \cite{Xu2007SurroundingGas} and Hao \cite{Hao2017EffectSurfaceRoughness} investigated the effects of surface roughness and the surrounding gas on the splashing and found that prompt splashing was due to surface roughness, but crown splashing was due to instabilities produced by the surrounding gas. As the environmental pressure is reduced, the corona splash is suppressed to a prompt splash, and on the further reduction of the pressure, prompt splash changes to lamella spread after a droplet impacting on a smooth moving substrate \cite{Hao2017SplashMovingSubstrate}. In addition, at reduced environmental pressure, Andrzej et al.\ \cite{Latka2012} found that the crown formation could be suppressed. Numerical simulations \cite{Guo2016DryWetAir} were conducted to investigate the effect of the environmental pressure on the splash after a drop impacting on a smooth surface, and the result showed that lowering the ambient gas density could suppress the splashing. Li et al.\ \cite{Li2017DoubleContact} and Lambley et al.\ \cite{Lambley2020SuperhydrophobicSurfacesEnvironmental} found that at reduced environmental pressure, the air disc entrapped under an impacting droplet became smaller than that at standard environmental pressure. When droplet impacts on the superhydrophobic substrate at reduced environmental pressure, the splashing can be eliminated, and the maximal diameter of the spreading lamella decreases with decreasing environmental pressure \cite{Tsai2011MicroscopicStructure, Tsai2010MicropatternAir}.

Besides the impact at reduced pressure, an important problem yet to be solved is the dynamics of droplet impact at high environmental pressure. The impact of droplets at increased pressure is not only a vital supplement to the previous studies at normal or reduced environmental pressure \cite{Xu2005Splashing}, but also crucial to many relevant applications such as fuel atomization in internal combustion engines \cite{Panao2005SprayImpingement}, spray cooling of high-power electronic devices, and droplet impact in nuclear power plants \cite{Crockett2010ErosionNuclear, Sanchez1988CorrosionErosion} at high pressure. For example, in nuclear power plants, the wall of the evaporator is unavoidable to be impacted by droplets in dispersed flow, which is an important open question for the safe operation of nuclear power plants. To ensure the safety of nuclear power plants, it is important not only to accurately describe the behavior of droplets impact but also to understand the dynamics of secondary droplets generated during the impact of droplets on the walls of the evaporator in such harsh environment of high pressure.

A droplet impacting on a substrate often splashes like a crown-shaped corona, breaking up into many tiny secondary droplets \cite{Bischofberger2016AirflowGenerated, Shen2016MoltenDroplet, Song2017VesicleSurfactant, Visser2015NumericalSimulation}. The splashing process is driven by the interplay of inertial, viscous, and capillary forces \cite{Worthington1876SymmetricSplashing, Mundo1995WallCollisions, Range1998SurfaceRoughness}. To quantify the effects of droplet sizes, droplet speeds, and the liquid properties on the crown propagation and splashing, the Weber number $\text{We} \equiv \rho {{U}^{2}}{{D}_{0}}/\sigma $ can be used to indicate the ratio between inertial and capillary forces, where $\rho $ and $\sigma$ are the density and surface tension of the liquid droplet, $U$ and ${{D}_{0}}$ denote the speed and the diameter of the droplet before the impact, respectively. The Reynolds number $\text{Re} \equiv \rho U{{D}_{0}}/\mu $ can be used to indicate the ratio of inertial and viscous forces, where $\mu $ is the dynamic viscosity of the liquid droplet. The Ohnesorge number $\text{Oh} \equiv {\sqrt{\text{We}}}/{\text{Re}}$ can be used to indicate the relationship between the inertial, capillary and viscous forces. The detailed mechanism of the ejection of secondary droplets from the crown and the effects of the impact parameters on the size of the secondary droplets are mainly studied at standard environmental pressure \cite{Wang2018SheetFragmentation, Zhang2010WavelengthSelection, Wang2018UnsteadyFragmentation, Wang2021UnsteadyFragmentation, Roisman2006FingeringSplashing, Villermaux2011DropFragmentation}. Different models have been proposed to explain the splashing mechanisms at standard environmental pressure, such as rim instability \cite{Wang2018SheetFragmentation, Wang2018UnsteadyFragmentation, Wang2021UnsteadyFragmentation, Roisman2006FingeringSplashing, Villermaux2011DropFragmentation}, eject sling-shot \cite{Thoroddsen2011SlingshotMechanism}, crown breakup \cite{Zhang2010WavelengthSelection}, and levitated viscous sheet \cite{Driscoll2010FilmFormation, Thoroddsen2010BubbleEntrapment}. Roisman et al.\ \cite{Roisman2006FingeringSplashing} studied the formation and growth of disturbance in the rim centerline by the transverse rim instability and found that the diameter of the fingers was similar to the size of the rim and also similar to the diameter of the outermost secondary droplets. The diameter of fingers, as it emerges from the base of the droplet, is governed by the viscous boundary layer along the substrate \cite{Thoroddsen2012Splashing}, and considering the unsteady nature of the flow, the diameter of the fingers can be calculated by ${{d}_{0}}/{{D}_{0}}\propto{{\text{Re}}^{-0.5}}$ \cite{Roisman2006FingeringSplashing, Thoroddsen2012Splashing, Xu2010InstabilityImpacting}.

For the impact of droplets on surfaces, there is a threshold from the spreading of the lamella to the splashing of the crown \cite{Driscoll2010FilmFormation, Latka2017ThresholdPressures, Latka2012}. The impact speed, the diameter, the surface tension, the viscosity, and the density of the liquid droplet have been known to be important for the splashing threshold \cite{Panao2005SprayImpingement, Xu2005Splashing, Stevens2014SplashingViscosity, Zhang2021ReversedViscositSplash}. Previous studies on the threshold of splashing have shown that the threshold is a function of the environmental pressure at reduced environmental pressure \cite{Latka2012, Li2017DoubleContact, Xu2007SurroundingGas, Stevens2014SplashingViscosity}. For example, silicone oil with different viscosities was used by Stevens et al.\ \cite{Stevens2014SplashingViscosity} to investigate the effect of the environmental pressure and the liquid viscosity on the splashing mechanism of droplets. They found that the threshold pressure of splashing decreased with increasing the viscosity at low viscosities of the droplet, whereas the trend was reversed at high viscosities. The effect of environmental pressure on the threshold for the transition from spreading to splashing was examined experimentally at increased environmental pressure \cite{Liu2010SplashingInclined}. Liu et al.\ found that the lamella spread horizontally along the surface at the standard environmental pressure and the spreading of the lamella was transformed to the crown splashing when the environmental pressure was increased to 2 bar after an FC-72 droplet impacting on a glass surface with the $\text{We} = 970$ \cite{Liu2010SplashingInclined}.

Since the environmental pressure can have a strong effect on the impact process, there is a great necessity to investigate the droplet impact dynamics at high environmental pressure and establish the relationship between the formation of secondary droplets and the environmental pressure. In this experimental study, we varied the environmental pressure in a wide range (from 0.1 to 7 bar) as well as other impact parameters. We found that the high environmental pressure can lead to the splashing of the crown in a `thread rupture' mode. The `thread rupture' is that during the growth of the liquid crown, the crown ruptures into liquid threads like stamens of flowers. The formation of the secondary droplets is from the tip of long liquid threads, which differentiates the breakup of droplets from the rim of the crown directly at the standard environmental pressure.

The rest of this paper is organized as follows. The experimental details are described in Section \ref{sec:sec02}. The results are discussed in Section \ref{sec:sec03}, including the dynamics of the impact process, the effect of the environmental pressure on the secondary droplets, the liquid crown, and the threshold of the splashing. Finally, conclusions are drawn in Section \ref{sec:sec04}.

\section{Experimental method }\label{sec:sec02}

\begin{figure}
\centering
  \includegraphics[width=0.9\columnwidth]{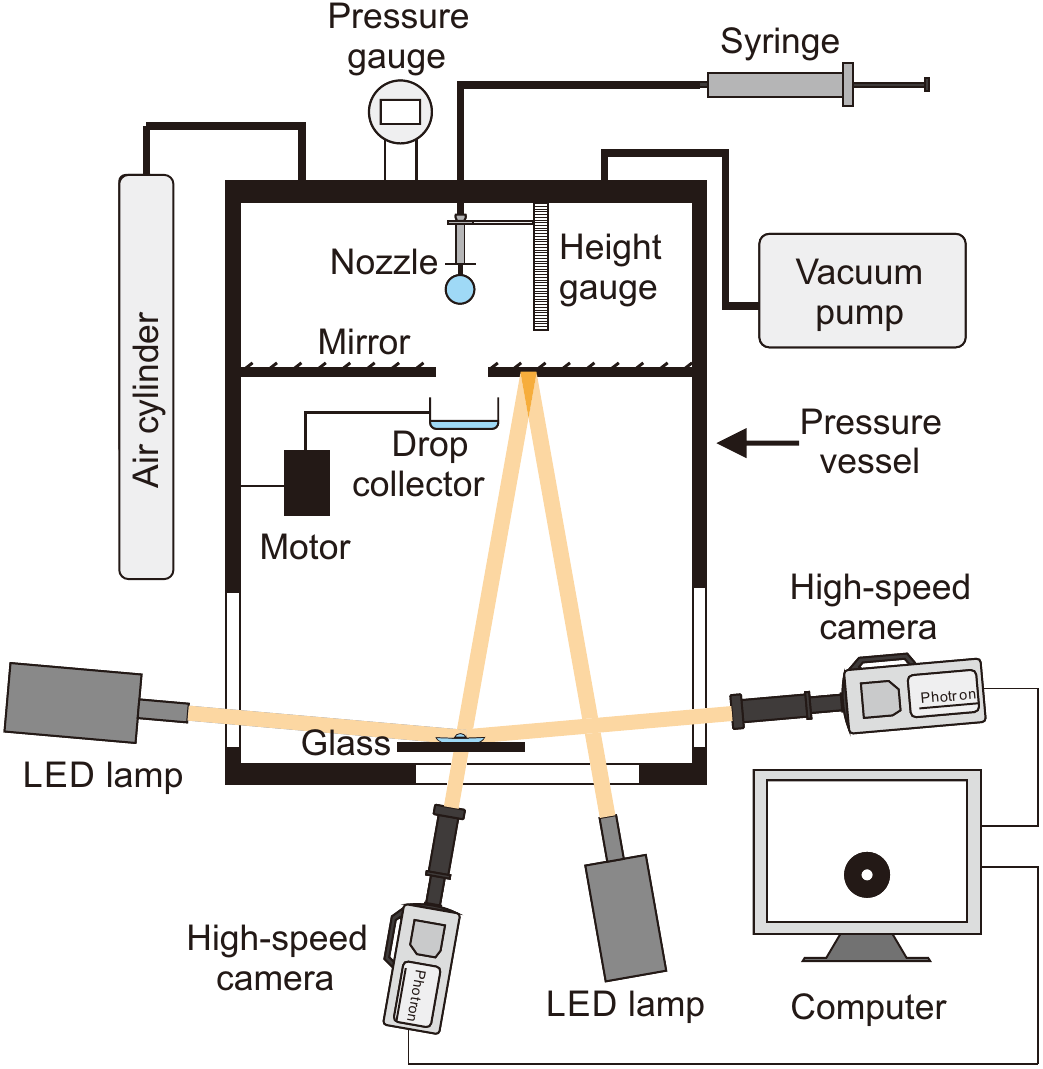}
  \caption{Schematic diagram of the experimental setup for the droplet impact at adjustable environmental pressure.}
  \label{fig:fig1}
\end{figure}

The impact was performed inside a pressure vessel as schematically described in Fig.\ \ref{fig:fig1}. The absolute pressure inside the pressure vessel, $P$, was varied from 0.1 to 7 bar by connecting to an air cylinder and a vacuum pump (see Table S1 in Supplementary Materials for variation of the air properties with pressure). Unless otherwise stated, the pressure value presented in this paper is the absolute pressure. Ethanol was used as the droplet fluid, and the density, the dynamic viscosity, and the surface tension are 789 kg/m$^3$, $1.2\times10^{-3}$ Pa$\cdot$s, and $22.3\times10^{-3}$ N/m, respectively. Ethanol droplets were released from a blunt syringe needle, which was connected to a syringe pushed by a syringe pump (Harvard Apparatus, Pump 11 elite Pico plus). The diameter of the droplets was varied between $2.0\pm 0.1$ and $3.5\pm 0.1$ mm by changing the size of the syringe needle. The speed of the droplets was varied by changing the height of the syringe needle from the substrate, and the size and the speed of the droplets were measured from the side-view images via an image processing algorithm using a customized Matlab program (see Supplementary Materials for the details). Dry smooth glass substrates were used as the target of impact, and they were laid horizontally inside the pressure vessel. The roughness of the glass surface is about 10 nm and the contact angle of ethanol on the surface is $7 \ \degree$ (measured from the side-view images). We used a fresh glass substrate in every impact event to avoid contamination by previous droplets. A droplet collector was used to screen the best droplets for impact and to avoid the uncontrolled continuous release of droplets. The processes of splashing and crown rupture were captured from the side-view and the bottom-view by using high-speed cameras (Photron Fastcam SA1.1). The side-view high-speed camera was at the speed of 16000 frames per second (fps) at the resolution of $\text{640}\times \text{512}$ pixels, corresponding to a spatial resolution of 22.2 $\mu$m/pixel, and the bottom-view high-speed camera was at the speed of 8000 frames per second (fps) at the resolution of $\text{1024}\times \text{752}$ pixels, corresponding to a spatial resolution of 12.5 $\mu$m/pixel. Due to the limited space in the pressure vessel, the side-view and the bottom view images were taken separately using repeated experiments by fixing the parameters of impact. The impact process was illuminated by high-power LED lights (Hecho S5000).

\section{Results and discussion}\label{sec:sec03}
\subsection{Impact morphology}\label{sec:sec031}

\begin{figure}
\centering
  \includegraphics[width=\columnwidth]{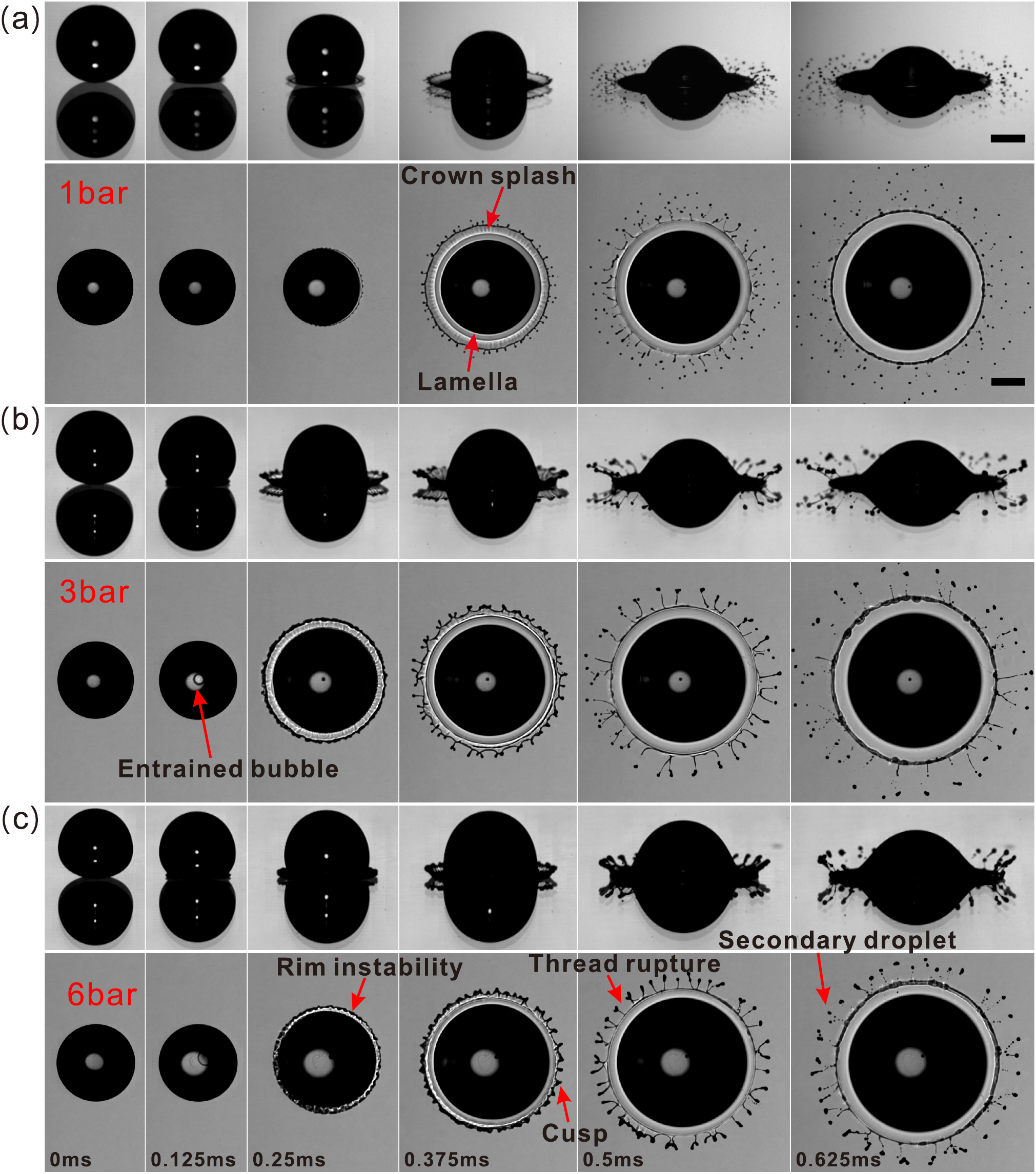}
  \caption{Image sequences of ethanol droplets impacting on glass surfaces with $\text{We} = 450$ at different environmental pressures. (a) 1 bar, (b) 3 bar, (c) 6 bar. The first and the second rows are side-view and bottom-view images. The scale bar is 1 mm. Multimedia view.}
  \label{fig:fig2}
\end{figure}

Figure \ref{fig:fig2} shows image sequences of ethanol droplets impacting on glass substrates with $\text{We} = 450$ at different environmental pressure. The first rows in Figs.\ \ref{fig:fig2}a, \ref{fig:fig2}b, and \ref{fig:fig2}c are the side-view images, and the second rows are the bottom-view images. A phenomenon can be observed that, as the environmental pressure increases, the rupture of the crown after the splashing is from liquid threads, as shown at 0.5 ms at the environmental pressure of 3 and 6 bar, which differentiates the breakup of the droplet from the rim of the crown directly in the standard pressure condition. Here in this section, we mainly consider the rupture of the crown in the splash regime, and the threshold of the environmental pressure in this splash regime will be further discussed in detail in Section \ref{sec:sec034}.

\begin{figure*}[tb]
\centering
  \includegraphics[width=1.4\columnwidth]{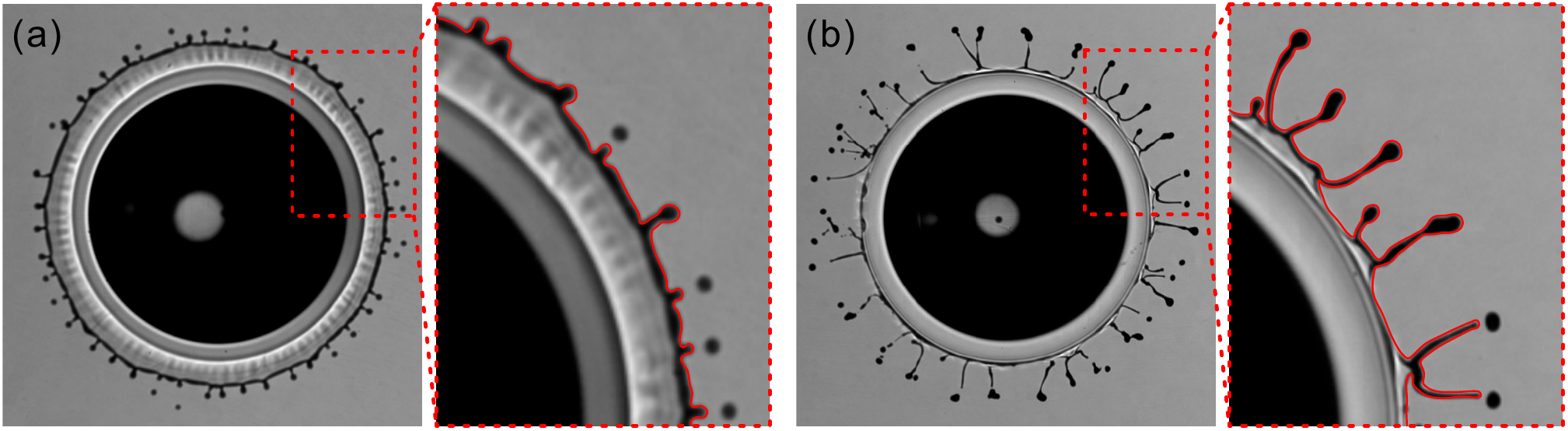}
  \caption{Comparison in the rupture of the crown between the droplet impact at standard environmental pressure and increased environmental pressure. (a) The finger jets at the rim of the liquid crown during the impact at standard environmental pressure. The finger jets are formed due to the development of the cusps on the rim of the liquid crown due to the instability of the rim. $\text{We} = 450$, ${P}=1\text{ bar}$. (b) Liquid threads connected to the lamella during the impact at increased environmental pressure. The liquid threads are formed after the rupture of the thin film of the liquid crown. $\text{We} = 450$, ${P}=3\text{ bar}$.}
  \label{fig:fig3}
\end{figure*}

\begin{figure*}[tb]
\centering
  \includegraphics[width=1.5\columnwidth]{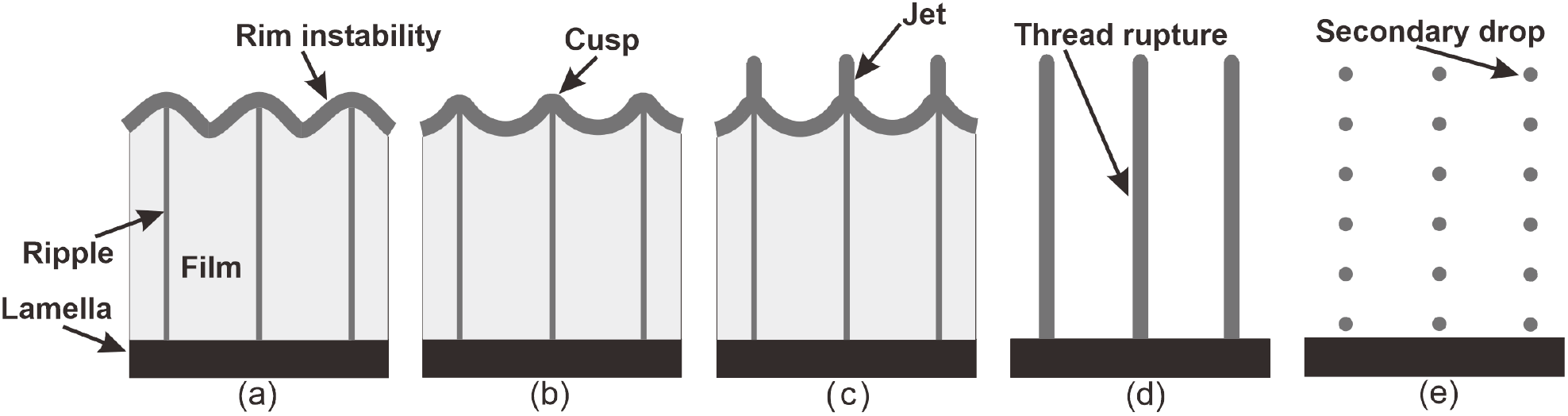}
  \caption{Schematic diagram of the process of crown rupture: (a) transverse rim instabilities on the film, (b) cusp formation, (c) generation of finger jets on the cusps, (d) thread rupture, and (e) generation of secondary droplets from crown rupture.}
  \label{fig:fig4}
\end{figure*}

The environmental pressure is vital in determining whether the rupture of the crown is from liquid threads. Figure \ref{fig:fig2}a (Multimedia view) shows the shape evolution of a droplet impacting on a substrate at the standard environmental pressure (1 bar). At 0.375 ms, the crown bends upwards from the substrate \cite{Xu2005Splashing}, and some tiny secondary droplets start to eject from the crown rim \cite{Xu2007SurroundingGas}. As we increase the environmental pressure from 1 to 3 bar, the main difference in the crown film is that the formation of the secondary droplets is from the tip of long liquid threads (see 0.5 ms in Fig.\ \ref{fig:fig2}b, Multimedia view), and this phenomenon is termed `thread rupture' to differentiate the breakup of the droplet from the rim of the crown directly at the standard environmental pressure. At 0.25 ms at the environmental pressure of 3 bar, the crown film has already bent up, as shown in Fig.\ \ref{fig:fig2}b, and in contrast, at 1 bar, only a weak disturbance occurs along the rim. In addition, the rim of the crown at 0.25 ms in Fig.\ \ref{fig:fig2}b is darker than the inside of the crown, which indicates that the rim of the crown is thicker than the inside of the crown, resulting in that the outer secondary droplets produced from the tip of the liquid threads are much larger than the inner secondary droplets after crown rupture, as shown at 0.625 ms in Fig.\ \ref{fig:fig2}b. As the environmental pressure is increased to 6 bar, many thick cusps are developed due to the rim instabilities on the liquid crown, as shown at 0.375 ms in Fig.\ \ref{fig:fig2}c. At the standard environmental pressure, cusps cannot develop into the liquid threads \cite{Agbaglah2013InstabilityRim, Roisman2006FingeringSplashing, Zhang2010WavelengthSelection} because of the detachment of secondary droplets from the rim of the liquid crown. In addition, the liquid threads at 6 bar are much shorter than that at 3 bar, as shown at 0.5 ms in Figs.\ \ref{fig:fig2}b and \ref{fig:fig2}c (Multimedia view), and this is because as the environmental pressure increases from 3 bar to 6 bar, the resistance of the surrounding air in the direction of the growth of the liquid crown increases due to the density of the surrounding air increases. The larger resistance of the surrounding air at 6 bar makes the length of the liquid crown shorter than the length of the liquid crown at 3 bar. Thus, the liquid threads at 6 bar are shorter than that at 3 bar after the rupture of the liquid crown. The secondary droplets at 6 bar are larger than that at 3 bar, as shown at 0.625 ms in Figs.\ \ref{fig:fig2}b and \ref{fig:fig2}c.

To highlight the difference between the thread rupture mode at increased environmental pressure and the breakup of the secondary droplet from the rim of the liquid crown directly at the standard environmental pressure, the shapes of the rims of the liquid crown are compared in Fig.\ \ref{fig:fig3}. At standard environmental pressure, the rim instability of the liquid crown, governed by the Rayleigh-Taylor instability and the Rayleigh-Plateau instability \cite{Rayleigh1878OnInstabilityJets, Taylor1950InstabilityLiquidSurfaces, Wang2021UnsteadyFragmentation, Zhang2010WavelengthSelection, Villermaux2011DropFragmentation}, develops into secondary droplets through the evolution of the finger jets \cite{Agbaglah2013InstabilityRim, Roisman2006FingeringSplashing, Yarin1995CapillaryWaves, Zhang2010WavelengthSelection}, as shown in Fig.\ \ref{fig:fig3}a. In contrast, at increased environmental pressure, due to the resistance of the surrounding air (with higher density than air at standard environmental pressure), the secondary droplet can hardly detach from the tip of the finger jet. In addition, due to the surface tension force of the rim, the rupture of the liquid crown introduces much fluid into the finger jets. As a consequence, long liquid threads are formed, as shown in Fig.\ \ref{fig:fig3}b. Therefore, finger jets and liquid threads are formed by different mechanisms, and the liquid threads at increased environmental pressure are much longer than the finger jets at the standard environmental pressure. Fig.\ \ref{fig:fig4} shows the rupture process of the liquid crown at increased environmental pressure in the `thread rupture' mode. The rupture process mainly includes the development of rim instabilities of the liquid crown, the formation of cusps, the generation of finger jets, the rupture of the liquid thread, and the formation of secondary droplets. At standard environmental pressure, the rim instability of the liquid crown develops into secondary droplets through the evolution of the cusps and finger jets \cite{Agbaglah2013InstabilityRim, Roisman2006FingeringSplashing, Yarin1995CapillaryWaves, Zhang2010WavelengthSelection}, but without noticeable liquid thread. The ripples of the liquid crown and the rupture of the liquid crown at increased environmental pressure are the main reason for the occurrence of the thread rupture, which will be further explained in Section \ref{sec:sec032}. Due to the resistance of the surrounding gas at high environmental pressure, the liquid crown quickly ruptures into long liquid threads. Finally, the threads further rupture into secondary droplets.

\subsection{Influence of environmental pressure on secondary droplets}\label{sec:sec032}

\begin{figure*}[t]
\centering
  \includegraphics[width=1.4\columnwidth]{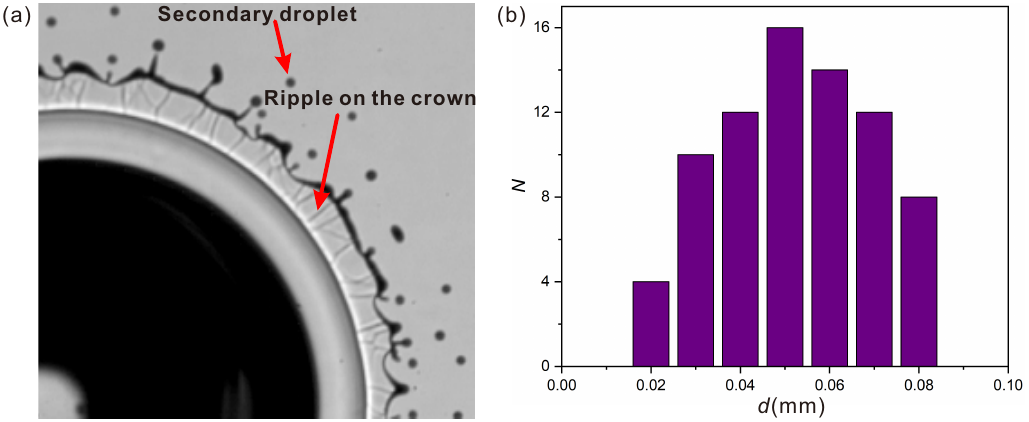}
  \caption{(a) Enlarged image of a droplet impacting with $\text{We} = 450$. There are many ripples on the crown film connecting the rim to the lamella. The rim is much thicker than the liquid crown, and secondary droplets are ejected from the rim of liquid crown. (b) Histograms of the sizes of the outermost secondary droplets for a droplet impacting on a glass surface with $\text{We} = 450$, ${P}=1\text{ bar}$. The histogram is an average over ten impact events under the same conditions.}
  \label{fig:fig5}
\end{figure*}

Figure \ref{fig:fig5}a shows a typical enlarged image of the crown splashing taken from the bottom view at $\text{We} = 450$. We can see that there are many ripples on the crown film, and the ripples are thicker than the nearby liquid crown. After the formation of the liquid crown, due to the resistance of the surrounding air at increased environmental pressure (high density of the surrounding air) in the direction of the growth of the liquid crown, and also due to the ripples on the liquid crown, the liquid crown quickly ruptures into long liquid threads. The outer secondary droplets produced from the rim are larger than the inner secondary droplets, as shown at 0.625 ms in Fig.\ \ref{fig:fig2}b. To quantitatively study the influence of the environmental pressure on the secondary droplets, we analyze the outermost secondary droplets ejected from the rim of the liquid crown or from the liquid threads at different environmental pressure. Figure \ref{fig:fig5}b shows the histogram of the sizes of the outermost secondary droplets for a droplet impacting on a glass surface with $\text{We} = 450$. The average diameter of the outermost secondary droplets is 0.052 mm. The sizes and the number of the secondary droplets were measured from the bottom-view images via an image processing algorithm using a customized Matlab program (see the Supplementary Materials for the details).

\begin{figure}[tb]
\centering
  \includegraphics[width=0.95\columnwidth]{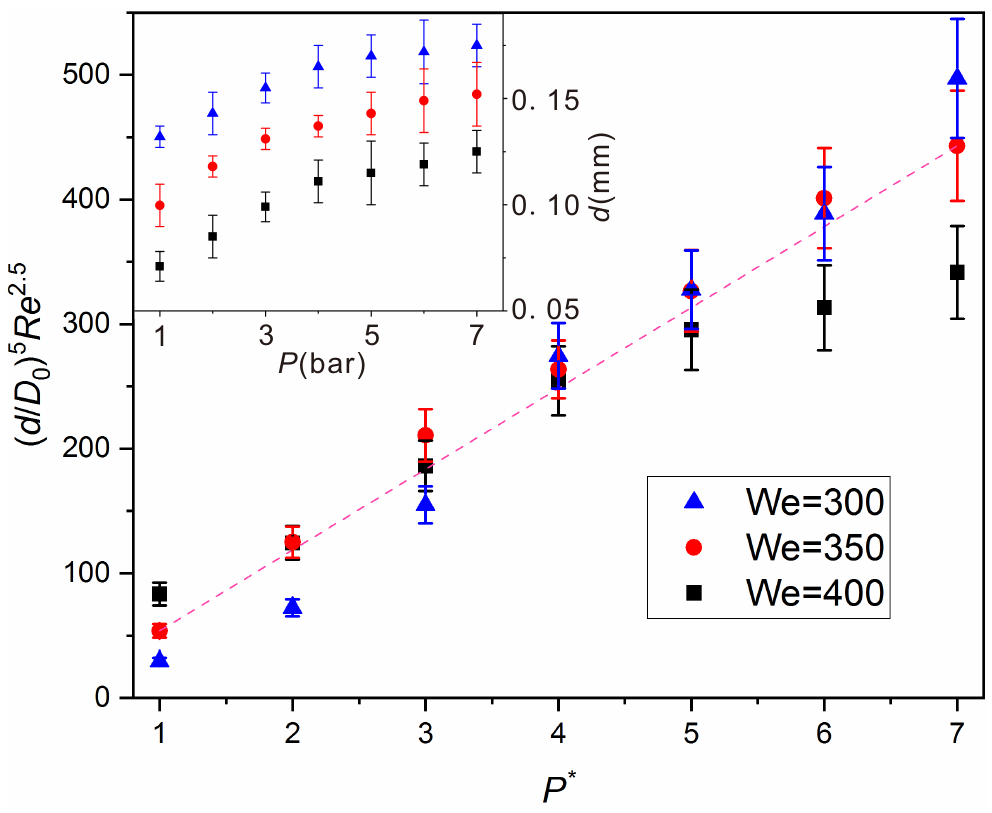}
  \caption{Effect of the environmental pressure on the scaled sizes of the secondary droplets produced from the crown rupture. The inset shows the sizes of secondary droplets versus environmental pressure in the dimensional form.}
  \label{fig:fig6}
\end{figure}

Figure \ref{fig:fig6} shows the influence of the environmental pressure on the secondary droplets produced from the crown rupture at different Weber numbers. The insets show the sizes of secondary droplets versus the environmental pressure. The main panels show the scaled values of secondary droplets as a function of the scaled environmental pressure. Please also refer to the Supplementary Materials for the size of secondary droplets for the impact of droplets with different droplet sizes. The measurement error of sizes of the secondary droplets depends on the resolution of the high-speed images. The resolution of the bottom-view images was $\text{1024}\times \text{752}$ pixels, corresponding to a spatial resolution of 12.5 $\mu$m/pixel. Considering the statistical nature of crown rupture, more than ten repeated impact events were measured under the same conditions. As shown in Fig.\ \ref{fig:fig6}, the size of the secondary droplets increases with the environmental pressure, and this trend can also be seen in the high-speed images at 0.625 ms in Fig.\ \ref{fig:fig2}. This trend is consistent with the results at reduced environmental pressure obtained by Lakta et al.\ \cite{Latka2012} From a fitting of the sizes of secondary droplets in the inset of Fig.\ \ref{fig:fig6}, the sizes of secondary droplets show an exponential dependence on the environmental pressure $d\propto{{P}^{\,0.2}}$, which can be written in dimensionless form $d/{{d}_{0}}\propto{(P/{{P}_{0}})}^{0.2}$, where ${{d}_{0}}$ is the average diameter of the secondary droplet at the standard pressure ${{P}_{0}}$. It has been reported that the sizes of secondary droplets at the standard pressure are scaled as ${{d}_{0}}/{{D}_{_{0}}}\propto{{\text{Re}}^{-0.5}}$, where $\text{Re} $ is the Reynolds number of droplet impact \cite{Roisman2006FingeringSplashing, Xu2010InstabilityImpacting} and $D_0$ is the initial diameter of the droplet before the impact. Since the density, the viscosity, and the surface tension of the ethanol droplet hardly change with the environmental pressure, by substituting ${{d}_{0}}$ into the formula at increased environmental pressures, the scaling of the secondary droplet size versus the environmental pressures can be described as ${{(d/{{D}_{0}})}^{5}}{{\text{Re}}^{2.5}}\propto P^*$, where $P^* \equiv P/{{P}_{0}}$. The experimental data of the diameter of the secondary droplet and the environmental pressure are plotted in the main panel of Fig.\ \ref{fig:fig6}.

\begin{figure*}[th]
\centering
  \includegraphics[width=2\columnwidth]{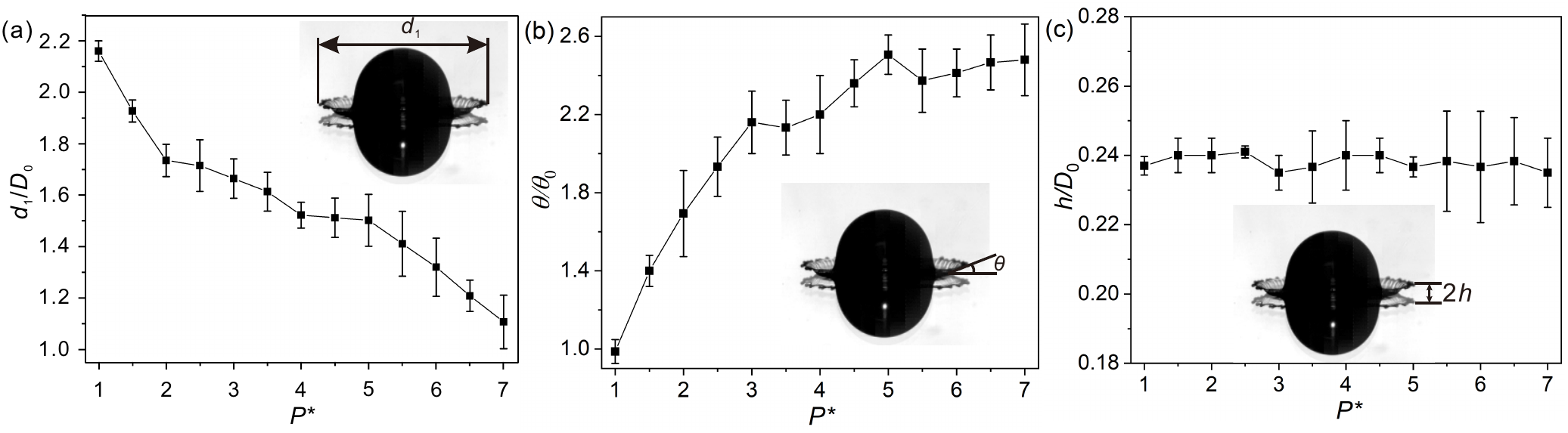}
  \caption{Effect of the environmental pressure on the crown splashing with $\text{We} = 400$. (a) The diameter of the liquid crown $d_1$ scaled by the initial diameter of the droplet $D_0$, ${D_0} = 2.2\pm0.1 \text{ mm}$. (b) The angle of the liquid crown $\theta$ scaled by the angle at the standard environmental pressure $\theta_0$, ${\theta_0} = 25\pm1 \degree$. (c) The height of the liquid crown $h$ scaled by the initial diameter of the droplet $D_0$.}
  \label{fig:fig7}
\end{figure*}

\subsection{Influence of environmental pressure on the liquid crown}\label{sec:sec033}

The rupture dynamics of the liquid crown is closely related to the morphology of liquid crown before the rupture. Therefore, in this section, the influence of the environmental pressure on the morphology of the liquid crown before the rupture is analyzed. The morphology of the liquid crown at different environmental pressure can be characterized mainly by three parameters: the diameter of the liquid crown rim, the angle of the liquid crown, and the height of the liquid crown, as sketched in the insets of Fig.\ \ref{fig:fig7}. These parameters are measured at the moment just before the rupture, and they are plotted against the environmental pressure in Fig.\ \ref{fig:fig7}.

The environmental pressure can suppress the movement of the liquid crown in the horizontal direction. As shown in Fig.\ \ref{fig:fig7}a, the diameter of the liquid crown decreases with increasing the environmental pressure. The angle of the liquid crown increases with the environmental pressure as shown in Fig.\ \ref{fig:fig7}b, and this is because the high environmental pressure can lift the liquid crown at larger angles from the substrate than that at the standard environmental pressure. The length of the liquid crown is expected to be larger, but the movement of the liquid crown is suppressed by the surrounding air with higher density in the vertical direction at increased environmental pressure. Therefore, the height of the liquid crown is almost constant at increased environmental pressure as shown in Fig.\ \ref{fig:fig7}c. The gas at high environmental pressure suppresses the increase of the height of the liquid crown, but the angle of the liquid crown increases, resulting in that the width of the rim of the liquid crown is shorter than that at the standard environmental pressure. As a consequence, the rim becomes thicker at higher environmental pressure and produces larger secondary droplets, as shown in Fig.\ \ref{fig:fig6}.

Previous researchers \cite{Latka2012, Stevens2014SplashingViscosity, Xu2007SurroundingGas, Xu2005Splashing} have shown that as the environmental pressure increased from vacuum to the standard environmental pressure, the liquid crown formed and became larger. However, our results show that if we continue to increase the environmental pressure from the standard environmental pressure, the liquid crown can be suppressed instead of becoming larger further, as shown in Fig.\ \ref{fig:fig7}a (see the Supplementary Materials). The diameter of the liquid crown decreases with increasing the environmental pressure. The reason is that the crown splashing is created by the air drag in the front of the lamella \cite{Xu2007SurroundingGas}, but after the formation of the liquid crown, the movement of the liquid crown can be suppressed by the surrounding air with higher density if we continue increasing the environmental pressure.

At increased environmental pressure, the length of the liquid thread decreases with increasing the environmental pressure at 0.5 ms in Figs.\ \ref{fig:fig2}b and \ref{fig:fig2}c. Since the rupture process is very quick, we can use the length of the liquid crown before the rupture to estimate the length of the liquid thread. As shown in Fig.\ \ref{fig:fig7}b, the angle of the liquid crown increases with the environmental pressure. The length of liquid thread can be estimated from the height and the angle of the liquid crown approximately ${{L}_{\text{thread}}}\approx h/\sin \theta $. The effect of the environmental pressure on the height of the liquid crown at the moment just before the rupture, $h$, is negligible, as shown in Fig.\ \ref{fig:fig7}c. Therefore, at increased environmental pressure, the length of the liquid thread decreases with increasing the environmental pressure, which can explain the decrease in the length of the liquid thread at 0.5 ms in Figs.\ \ref{fig:fig2}b and \ref{fig:fig2}c.

\subsection{Influence of environmental pressure on the threshold of splashing}\label{sec:sec034}

\begin{figure}[tb]
\centering
  \includegraphics[width=\columnwidth]{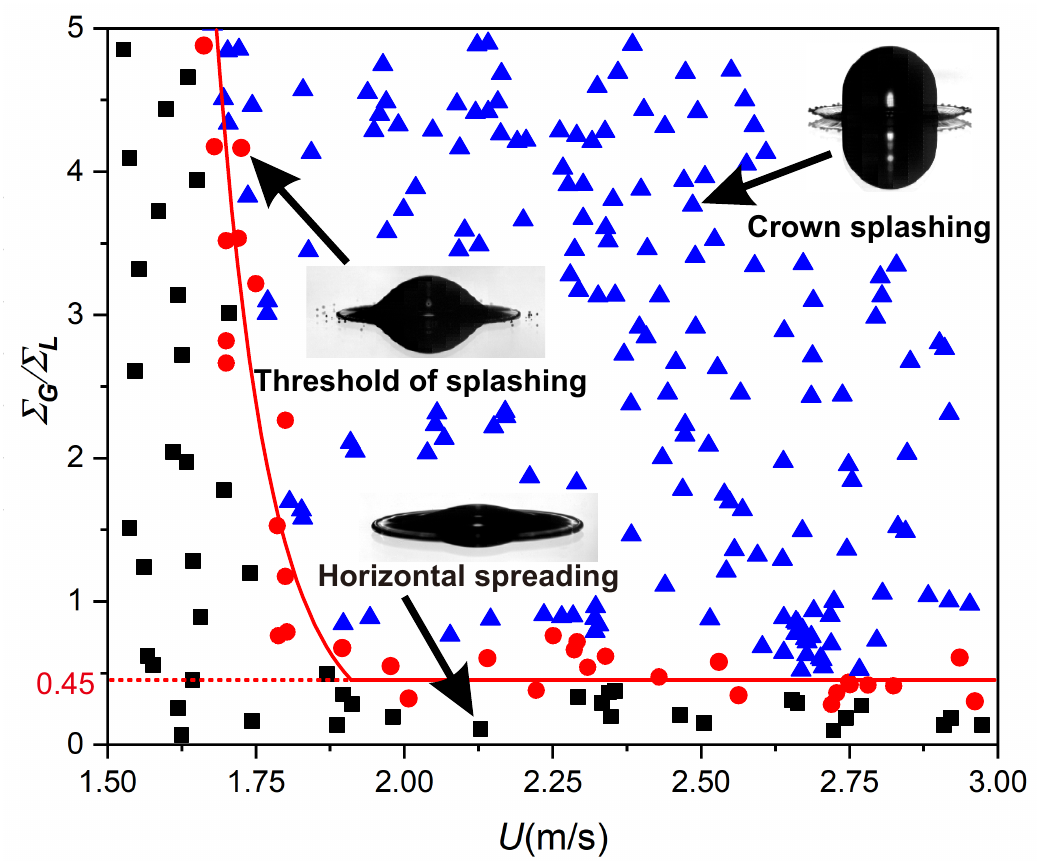}
  \caption{Regime map of ${{\Sigma }_{G}}/{{\Sigma }_{L}}$ versus the speed of droplet impact at different environmental pressure. The red dot shows the threshold of splashing, and the horizontal spreading is indicated by the solid black square, and the crown splashing is indicated by the solid blue triangle. The solid red line is the approximate threshold of splashing, obtained according to the threshold point. When droplets impact at reduced environmental pressure,  ${{\Sigma }_{G}}/{{\Sigma }_{L}}$ at the threshold is almost a constant 0.45. However, at increased environmental pressure, ${{\Sigma }_{G}}/{{\Sigma }_{L}}$ at the threshold decreases with increasing the speed of droplet impact.}
  \label{fig:fig8}
\end{figure}

Upon the impact of droplets on solid surfaces, the splashing phenomenon occurs only when the inertia is sufficiently large and there is a threshold from the spreading of the lamella to the splashing of the crown. In this section, the threshold of splashing at increased environmental pressure is analyzed. We firstly consider the widely used theory based on the ``water hammer'' effect to analyze the effect of the environmental pressure on the threshold of splashing proposed by Xu et al \cite{Xu2005Splashing}. When the ratio between the destabilizing stress due to the gas ${{\Sigma }_{G}}$ and the stabilizing stress due to surface tension ${{\Sigma }_{L}}$ (i.e., ${{{\Sigma }_{G}}}/{{{\Sigma }_{L}}}\text{=}\sqrt{\gamma {{M}_{g}}}p\sqrt{\frac{{{D}_{0}}U}{4{{k}_{B}}T}}\frac{{{\nu }_{l}}}{\sigma }$
where $\gamma $ is the adiabatic constant of the gas, ${{M}_{g}}$ is the molecular weight of the gas, ${{k}_{B}}$ is the Boltzmann constant, $T$ is the gas temperature, ${{\nu }_{l}}$ is the kinematic viscosity of the liquid) is greater than the critical number 0.45, the crown splashing occurs. The regime map of ${{\Sigma }_{G}}/{{\Sigma }_{L}}$ versus the speed of droplet impact at different environmental pressure is shown in Fig.\ \ref{fig:fig8}. A threshold scenario of droplet impact is obtained, as shown in the inset images in Fig.\ \ref{fig:fig8}. In the spreading condition, the lamella spreads horizontally along the substrate without any liftup. In contrast, in the splashing condition, the lamella lifts up from the substrate. In the threshold scenario shown in Fig.\ \ref{fig:fig8} (the inset image labeled with `Threshold of splashing'), the secondary droplets emit from the lamella directly without any liftup of the lamella. The criterion by Xu et al. \cite{Xu2005Splashing} is only applicable to reduced environmental pressure, but cannot predict the threshold of splashing at increased environmental pressure, as shown in the regime of impact in Fig.\ \ref{fig:fig8}.

As the criterion by Xu et al.\ \cite{Xu2005Splashing} cannot predict the threshold of splashing at increased environmental pressure, we consider the criterion proposed by Riboux and Gordillo \cite{Riboux2014CriticalSpeed} for splashing based on the aerodynamics analysis in which the resistance from the air ahead drives the lamella to detach from the substrate. The criterion applies to the low viscosity liquids (${{\text{Re}}^{1/6}}\text{O}{{\text{h}}^{2/3}}<0.22$) and the millimeter droplets and was later demonstrated for reduced pressure \cite{Ashida2020PromptSplashing, Gordillo2019AerodynamicSplashing, Hao2019InclinedSurface}. In this criterion, the expression of the droplet speed before the impact, $K_l$, is used to predict the threshold of splashing. In the limit $\text{W}{{\text{e}}_{\lambda }}> 3{{\left( {{\mu }_{g}}/\mu  \right)}^{3/4}}\text{O}{{\text{h}}^{1/4}}$, where $\lambda $ is the mean free path of the gas molecules, ${{\mu }_{g}}$ is the viscosity of the gas, and $\text{We}_{\lambda } \equiv \text{We}{\lambda }/{{{D}_{0}}}$, the threshold of splashing can be expressed as
\begin{equation}\label{eq:eq01}
  {U={K}_{l}}\sim\ln \left[ \frac{1+{{K}_{a}}\text{W}{{\text{e}}^{1/12}}\text{O}{{\text{h}}^{1/2}}}{C\text{W}{{\text{e}}_{\lambda }}}\left( 1+C\text{W}{{\text{e}}_{\lambda }} \right) \right]
\end{equation}
where ${{K}_{a}}$ is a proportionality constant ${{K}_{a}}=0.7$ and $C$ is a fitting constant $C=19$ \cite{Gordillo2019AerodynamicSplashing}. The mean free path of gas molecules follows
\begin{equation}\label{eq:eq02}
  \lambda \text{=}{{\lambda }_{0}}\frac{T}{{{T}_{0}}}\frac{{{P}_{0}}}{P}
\end{equation}
where ${{\lambda }_{0}}=6.5\times {{10}^{- 8}}\text{m}$ is the mean free path of gas molecules at ${{T}_{0}}=300\,\text{K}$ and ${{P}_{0}}=1\text{ bar}$. In addition, in the limit $\text{W}{{\text{e}}_{\lambda }}<3{{\left( {{\mu }_{g}}/\mu  \right)}^{3/4}}\text{O}{{\text{h}}^{1/4}}$, the threshold of splashing can be expressed as
\begin{equation}\label{eq:eq03}
  {U={K}_{l}}\sim\ln \left( A{{\ell }_{\mu }}{{\ell }_{g}} \right)
\end{equation}
where ${{\ell }_{g}}$ is the slip length of liquid at the substrate under the wedge of the lamella ${{\ell }_{g}}\text{=W}{{\text{e}}_{\lambda }}{{(1+{{K}_{a}}\text{W}{{\text{e}}^{1/12}}\text{O}{{\text{h}}^{1/2}})}^{-1}}$, ${{\ell }_{\mu }}$ is the slip length at the gas-liquid interface under the wedge of the lamella ${{\ell }_{\mu }} = {{\ell }_{g}}\text{+}{{\left( {\mu }/{{{\mu }_{g}}} \right)}^{{-3/4}}}\text{O}{{\text{h}}^{1/4}}$, $A$ is a fitting constant. In the derivation of Ref. \cite{Gordillo2019AerodynamicSplashing}, the expression of ${{\ell }_{\mu }}$ is simplified to ${{\ell }_{\mu }}\text{=}{{\left( {\mu }/{{{\mu }_{g}}} \right)}^{-3/4}}{{\text{Oh}}^{1/4}}$ by neglecting ${{\ell }_{g}}$. However, at different environmental pressure varied from 0.1 to 7 bar, ${{\ell }_{g}}$ is an important contribution to ${{\ell }_{\mu }}$ and cannot be neglected because ${{\ell }_{g}}$ depends on the mean free path of the gas molecules $\lambda $, which depends on the environmental pressure as described by Eq.\ (\ref{eq:eq02}). For example, at the threshold of splashing, when the environmental pressure is 0.4 bar, ${{\ell }_{g}}/{{\ell }_{\mu }}\approx 0.9$, and when the environmental pressure is 1 bar, ${{\ell }_{g}}/{{\ell }_{\mu }}\approx 0.6$. Therefore, ${{\ell }_{g}}$ is an important contribution to ${{\ell }_{\mu }}$, and its contribution should be considered.

\begin{figure}[tb]
\centering
  \includegraphics[width=\columnwidth]{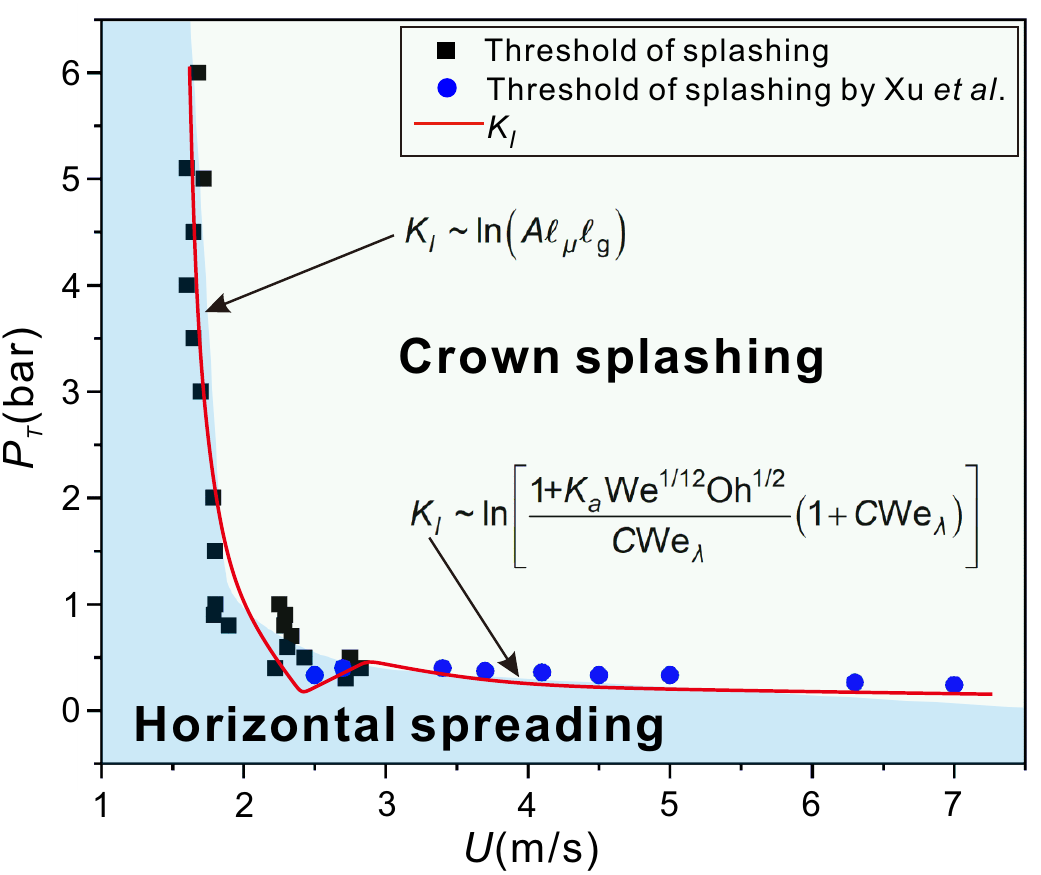}
  \caption{Threshold pressure versus the speed of droplet impact. Below the threshold of splashing, the horizontal spreading of droplets occurs, while above the threshold, the crown splashing of droplets occurs. The black squares indicate the threshold data of splashing from this study, and the blue dots indicate the threshold data from Xu et al.\ \cite{Xu2005Splashing} The lines indicate the thresholds in Eqs.\ (\ref{eq:eq01}) and (\ref{eq:eq03}) where the contribution of ${{\ell }_{g}}$ to ${{\ell }_{\mu }}$ is considered. The fitting constant $A$ is $A=6.6\times 10^{{-7}}$.}
  \label{fig:fig9}
\end{figure}

The threshold speed of droplet impact for splashing, dependent on environmental pressure, is drawn in Fig.\ \ref{fig:fig9}. When the speed of impact is greater than 3 m/s, the threshold pressure decreases slowly with increasing the speed of impact. However, when the speed of impact is less than 2.5 m/s, the threshold pressure decreases quickly with increasing the speed of impact. There is a non-monotonic interval of the threshold pressure of splashing versus the speed of droplet impact as the environmental pressure increases, which is consistent with previous studies \cite{Gordillo2019AerodynamicSplashing, Riboux2014CriticalSpeed, Xu2005Splashing, Hao2019InclinedSurface}. Please also refer to the Supplementary Materials for the effect of the droplet size on the threshold pressure of splashing. At the threshold of splashing, the contribution of ${{\ell }_{g}}$ to ${{\ell }_{\mu }}$ decreases with increasing the environmental pressure. Since ${{\ell }_{g}}$ and ${{\ell }_{\mu }}$ decreases with increasing the environmental pressure, the threshold speed of splashing $U$ (calculated by the ${{K}_{l}}$ in Eqs.\ (\ref{eq:eq01}) and (\ref{eq:eq03})) decreases with increasing the environmental pressure, but the slope of the $P_T$ versus $U$ increases with increasing the environmental pressure, as shown in Fig.\ \ref{fig:fig9}. At increased environmental pressure, the wedge of the lamella is prevented from moving forward and is driven to detach from the substrate by the air ahead, which has a higher density than that at standard environmental pressure. Therefore, the liquid crown can splash at a lower impact speed of the impact at higher environmental pressure, as shown in Fig.\ \ref{fig:fig9}.

\section{Conclusions}\label{sec:sec04}
This paper experimentally investigates the impact of droplets on dry smooth surfaces by using high-speed photography at increased environmental pressure. The effects of the environmental pressure on the splashing and rupture of the crown during the impact process are analyzed. With increasing the environmental pressure, a phenomenon of thread rupture occurs during the crown splashing. The formation of secondary droplets from the crown splashing includes several stages, namely rim instability, cusps along the rim, jets, thread rupture, and finally the formation of secondary droplets. The sizes of the secondary droplets are measured from the high-speed images, and the results show that high environmental pressure can promote the crown rupture, and the size of the secondary droplets from the rim of the liquid crown increases with the environmental pressure. The threshold for the transition from spreading to splashing during the impact is obtained from our experimental data, and is analyzed based on the theory of aerodynamics analysis of the lamella. At increased environmental pressure, the surrounding gas plays an important role on the threshold of the splashing because the wedge of the lamella is prevented from moving forward and is driven to detach from the substrate by the air ahead, which has a higher density due to the higher pressure. The threshold speed of the impact decreases with increasing the environmental pressure, and the slope of the threshold pressure versus the impact speed increases with increasing the environmental pressure.

This paper studies droplets impacting at increased environmental pressure. There are still many open questions for the impact process at the increased environmental pressure, such as the development of the azimuthal instability of the rim, the pinch-off of secondary droplets from the rim, and the liftoff of the lamella. The study of droplet impact at increased environmental pressure will not only deepen our understanding of the impact dynamics, but also will be useful for the optimization of the relevant applications.

\section*{Supplementary Materials}
See the Supplementary Materials for the properties of the air at different environmental pressures (Sec.\ S1), the details of image processing (Sec.\ S2), the effects of the droplet size on the threshold pressure of splashing and the size of the secondary droplet (Sec.\ S3), and the threshold pressure from the splash enhancement to the suppression of liquid crown (Sec.\ S4).

\section*{Acknowledgements}
This work is supported by the National Natural Science Foundation of China (Grant nos.\ 51676137, 52176083, and 51920105010).

\section*{Conflict of interest}
The authors have no conflicts to disclose.

\section*{Data Availability Statement}
The data that support the findings of this study are available from the corresponding author upon reasonable request.

\bibliography{HighPressureImpact}

\begin{thebibliography}{47}%
\makeatletter
\providecommand \@ifxundefined [1]{%
 \@ifx{#1\undefined}
}%
\providecommand \@ifnum [1]{%
 \ifnum #1\expandafter \@firstoftwo
 \else \expandafter \@secondoftwo
 \fi
}%
\providecommand \@ifx [1]{%
 \ifx #1\expandafter \@firstoftwo
 \else \expandafter \@secondoftwo
 \fi
}%
\providecommand \natexlab [1]{#1}%
\providecommand \enquote  [1]{``#1''}%
\providecommand \bibnamefont  [1]{#1}%
\providecommand \bibfnamefont [1]{#1}%
\providecommand \citenamefont [1]{#1}%
\providecommand \href@noop [0]{\@secondoftwo}%
\providecommand \href [0]{\begingroup \@sanitize@url \@href}%
\providecommand \@href[1]{\@@startlink{#1}\@@href}%
\providecommand \@@href[1]{\endgroup#1\@@endlink}%
\providecommand \@sanitize@url [0]{\catcode `\\12\catcode `\$12\catcode
  `\&12\catcode `\#12\catcode `\^12\catcode `\_12\catcode `\%12\relax}%
\providecommand \@@startlink[1]{}%
\providecommand \@@endlink[0]{}%
\providecommand \url  [0]{\begingroup\@sanitize@url \@url }%
\providecommand \@url [1]{\endgroup\@href {#1}{\urlprefix }}%
\providecommand \urlprefix  [0]{URL }%
\providecommand \Eprint [0]{\href }%
\providecommand \doibase [0]{http://dx.doi.org/}%
\providecommand \selectlanguage [0]{\@gobble}%
\providecommand \bibinfo  [0]{\@secondoftwo}%
\providecommand \bibfield  [0]{\@secondoftwo}%
\providecommand \translation [1]{[#1]}%
\providecommand \BibitemOpen [0]{}%
\providecommand \bibitemStop [0]{}%
\providecommand \bibitemNoStop [0]{.\EOS\space}%
\providecommand \EOS [0]{\spacefactor3000\relax}%
\providecommand \BibitemShut  [1]{\csname bibitem#1\endcsname}%
\let\auto@bib@innerbib\@empty
\bibitem [{\citenamefont {Wang}, \citenamefont {Wilkinson},\ and\ \citenamefont
  {Drallmeier}(2004)}]{Wang2004FilmBreakup}%
  \BibitemOpen
  \bibfield  {author} {\bibinfo {author} {\bibfnamefont {Y.~P.}\ \bibnamefont
  {Wang}}, \bibinfo {author} {\bibfnamefont {G.~B.}\ \bibnamefont {Wilkinson}},
  \ and\ \bibinfo {author} {\bibfnamefont {J.~A.}\ \bibnamefont {Drallmeier}},\
  }\bibfield  {title} {\enquote {\bibinfo {title} {Parametric study on the fuel
  film breakup of a cold start {PFI} engine},}\ }\href {\doibase
  10.1007/S00348-004-0827-x} {\bibfield  {journal} {\bibinfo  {journal}
  {Experiments in Fluids}\ }\textbf {\bibinfo {volume} {37}},\ \bibinfo {pages}
  {385--398} (\bibinfo {year} {2004})}\BibitemShut {NoStop}%
\bibitem [{\citenamefont {Bergeron}\ \emph {et~al.}(2000)\citenamefont
  {Bergeron}, \citenamefont {Bonn}, \citenamefont {Martin},\ and\ \citenamefont
  {Vovelle}}]{Bergeron2000NatureImpactPolymer}%
  \BibitemOpen
  \bibfield  {author} {\bibinfo {author} {\bibfnamefont {V.}~\bibnamefont
  {Bergeron}}, \bibinfo {author} {\bibfnamefont {D.}~\bibnamefont {Bonn}},
  \bibinfo {author} {\bibfnamefont {J.~Y.}\ \bibnamefont {Martin}}, \ and\
  \bibinfo {author} {\bibfnamefont {L.}~\bibnamefont {Vovelle}},\ }\bibfield
  {title} {\enquote {\bibinfo {title} {Controlling droplet deposition with
  polymer additives},}\ }\href {\doibase 10.1038/35015525} {\bibfield
  {journal} {\bibinfo  {journal} {Nature}\ }\textbf {\bibinfo {volume} {405}},\
  \bibinfo {pages} {772--775} (\bibinfo {year} {2000})}\BibitemShut {NoStop}%
\bibitem [{\citenamefont {Yarin}(2006)}]{Yarin2006RecoilBouncing}%
  \BibitemOpen
  \bibfield  {author} {\bibinfo {author} {\bibfnamefont {A.~L.}\ \bibnamefont
  {Yarin}},\ }\bibfield  {title} {\enquote {\bibinfo {title} {Drop impact
  dynamics: splashing, spreading, receding, bouncing...}}\ }\href {\doibase
  10.1146/annurev.fluid.38.050304.092144} {\bibfield  {journal} {\bibinfo
  {journal} {Annual Review of Fluid Mechanics}\ }\textbf {\bibinfo {volume}
  {38}},\ \bibinfo {pages} {159--192} (\bibinfo {year} {2006})}\BibitemShut
  {NoStop}%
\bibitem [{\citenamefont {Wirth}\ and\ \citenamefont
  {Storp}(1991)}]{Wirth1991AgricultureSpray}%
  \BibitemOpen
  \bibfield  {author} {\bibinfo {author} {\bibfnamefont {W.}~\bibnamefont
  {Wirth}}\ and\ \bibinfo {author} {\bibfnamefont {S.}~\bibnamefont {Storp}},\
  }\bibfield  {title} {\enquote {\bibinfo {title} {Mechanisms controlling leaf
  retention of agricultural spray solutions},}\ }\href {\doibase
  10.1002/ps.2780330403} {\bibfield  {journal} {\bibinfo  {journal} {Pesticide
  Science}\ }\textbf {\bibinfo {volume} {33}},\ \bibinfo {pages} {411--421}
  (\bibinfo {year} {1991})}\BibitemShut {NoStop}%
\bibitem [{\citenamefont {Xu}, \citenamefont {Zhang},\ and\ \citenamefont
  {Nagel}(2005)}]{Xu2005Splashing}%
  \BibitemOpen
  \bibfield  {author} {\bibinfo {author} {\bibfnamefont {L.}~\bibnamefont
  {Xu}}, \bibinfo {author} {\bibfnamefont {W.~W.}\ \bibnamefont {Zhang}}, \
  and\ \bibinfo {author} {\bibfnamefont {S.~R.}\ \bibnamefont {Nagel}},\
  }\bibfield  {title} {\enquote {\bibinfo {title} {Drop splashing on a dry
  smooth surface},}\ }\href {\doibase 10.1103/PhysRevLett.94.184505} {\bibfield
   {journal} {\bibinfo  {journal} {Physical Review Letters}\ }\textbf {\bibinfo
  {volume} {94}},\ \bibinfo {pages} {184505} (\bibinfo {year}
  {2005})}\BibitemShut {NoStop}%
\bibitem [{\citenamefont {Xu}, \citenamefont {Barcos},\ and\ \citenamefont
  {Nagel}(2007)}]{Xu2007SurroundingGas}%
  \BibitemOpen
  \bibfield  {author} {\bibinfo {author} {\bibfnamefont {L.}~\bibnamefont
  {Xu}}, \bibinfo {author} {\bibfnamefont {L.}~\bibnamefont {Barcos}}, \ and\
  \bibinfo {author} {\bibfnamefont {S.~R.}\ \bibnamefont {Nagel}},\ }\bibfield
  {title} {\enquote {\bibinfo {title} {Splashing of liquids: interplay of
  surface roughness with surrounding gas},}\ }\href {\doibase
  10.1103/PhysRevE.76.066311} {\bibfield  {journal} {\bibinfo  {journal}
  {Physical Review E}\ }\textbf {\bibinfo {volume} {76}},\ \bibinfo {pages}
  {066311} (\bibinfo {year} {2007})}\BibitemShut {NoStop}%
\bibitem [{\citenamefont {Latka}\ \emph {et~al.}(2012)\citenamefont {Latka},
  \citenamefont {Strandburg-Peshkin}, \citenamefont {Driscoll}, \citenamefont
  {Stevens},\ and\ \citenamefont {Nagel}}]{Latka2012}%
  \BibitemOpen
  \bibfield  {author} {\bibinfo {author} {\bibfnamefont {A.}~\bibnamefont
  {Latka}}, \bibinfo {author} {\bibfnamefont {A.}~\bibnamefont
  {Strandburg-Peshkin}}, \bibinfo {author} {\bibfnamefont {M.~M.}\ \bibnamefont
  {Driscoll}}, \bibinfo {author} {\bibfnamefont {C.~S.}\ \bibnamefont
  {Stevens}}, \ and\ \bibinfo {author} {\bibfnamefont {S.~R.}\ \bibnamefont
  {Nagel}},\ }\bibfield  {title} {\enquote {\bibinfo {title} {Creation of
  prompt and thin-sheet splashing by varying surface roughness or increasing
  air pressure},}\ }\href {\doibase 10.1103/PhysRevLett.109.054501} {\bibfield
  {journal} {\bibinfo  {journal} {Physical Review Letters}\ }\textbf {\bibinfo
  {volume} {109}},\ \bibinfo {pages} {054501} (\bibinfo {year}
  {2012})}\BibitemShut {NoStop}%
\bibitem [{\citenamefont {Li}\ \emph {et~al.}(2017)\citenamefont {Li},
  \citenamefont {Langley}, \citenamefont {Tian}, \citenamefont {Hicks},\ and\
  \citenamefont {Thoroddsen}}]{Li2017DoubleContact}%
  \BibitemOpen
  \bibfield  {author} {\bibinfo {author} {\bibfnamefont {E.~Q.}\ \bibnamefont
  {Li}}, \bibinfo {author} {\bibfnamefont {K.~R.}\ \bibnamefont {Langley}},
  \bibinfo {author} {\bibfnamefont {Y.~S.}\ \bibnamefont {Tian}}, \bibinfo
  {author} {\bibfnamefont {P.~D.}\ \bibnamefont {Hicks}}, \ and\ \bibinfo
  {author} {\bibfnamefont {S.~T.}\ \bibnamefont {Thoroddsen}},\ }\bibfield
  {title} {\enquote {\bibinfo {title} {Double contact during drop impact on a
  solid under reduced air pressure},}\ }\href {\doibase
  10.1103/PhysRevLett.119.214502} {\bibfield  {journal} {\bibinfo  {journal}
  {Physical Review Letters}\ }\textbf {\bibinfo {volume} {119}},\ \bibinfo
  {pages} {214502} (\bibinfo {year} {2017})}\BibitemShut {NoStop}%
\bibitem [{\citenamefont {Latka}(2017)}]{Latka2017ThresholdPressures}%
  \BibitemOpen
  \bibfield  {author} {\bibinfo {author} {\bibfnamefont {A.}~\bibnamefont
  {Latka}},\ }\bibfield  {title} {\enquote {\bibinfo {title} {Thin-sheet
  creation and threshold pressures in drop splashing},}\ }\href {\doibase
  10.1039/c6sm02321e} {\bibfield  {journal} {\bibinfo  {journal} {Soft Matter}\
  }\textbf {\bibinfo {volume} {13}},\ \bibinfo {pages} {740--747} (\bibinfo
  {year} {2017})}\BibitemShut {NoStop}%
\bibitem [{\citenamefont {Hao}(2017)}]{Hao2017EffectSurfaceRoughness}%
  \BibitemOpen
  \bibfield  {author} {\bibinfo {author} {\bibfnamefont {J.~G.}\ \bibnamefont
  {Hao}},\ }\bibfield  {title} {\enquote {\bibinfo {title} {Effect of surface
  roughness on droplet splashing},}\ }\href {\doibase 10.1063/1.5005990}
  {\bibfield  {journal} {\bibinfo  {journal} {Physics of Fluids}\ }\textbf
  {\bibinfo {volume} {12}},\ \bibinfo {pages} {122105} (\bibinfo {year}
  {2017})}\BibitemShut {NoStop}%
\bibitem [{\citenamefont {Hao}\ and\ \citenamefont
  {Green}(2017)}]{Hao2017SplashMovingSubstrate}%
  \BibitemOpen
  \bibfield  {author} {\bibinfo {author} {\bibfnamefont {J.~G.}\ \bibnamefont
  {Hao}}\ and\ \bibinfo {author} {\bibfnamefont {S.~I.}\ \bibnamefont
  {Green}},\ }\bibfield  {title} {\enquote {\bibinfo {title} {Splash threshold
  of a droplet impacting a moving substrate},}\ }\href {\doibase
  10.1063/1.4972976} {\bibfield  {journal} {\bibinfo  {journal} {Physics of
  Fluids}\ }\textbf {\bibinfo {volume} {29}},\ \bibinfo {pages} {012103}
  (\bibinfo {year} {2017})}\BibitemShut {NoStop}%
\bibitem [{\citenamefont {Guo}\ and\ \citenamefont
  {Sussman}(2016)}]{Guo2016DryWetAir}%
  \BibitemOpen
  \bibfield  {author} {\bibinfo {author} {\bibfnamefont {Y.~S.}\ \bibnamefont
  {Guo}, \bibfnamefont {Y.~S.~Lian}}\ and\ \bibinfo {author} {\bibfnamefont
  {M.}~\bibnamefont {Sussman}},\ }\bibfield  {title} {\enquote {\bibinfo
  {title} {Investigation of drop impact on dry and wet surfaces with
  consideration of surrounding air},}\ }\href {\doibase 10.1063/1.4958694}
  {\bibfield  {journal} {\bibinfo  {journal} {Physics of Fluids}\ }\textbf
  {\bibinfo {volume} {28}},\ \bibinfo {pages} {073303} (\bibinfo {year}
  {2016})}\BibitemShut {NoStop}%
\bibitem [{\citenamefont {Lambley}\ and\ \citenamefont
  {Poulikakos}(2020)}]{Lambley2020SuperhydrophobicSurfacesEnvironmental}%
  \BibitemOpen
  \bibfield  {author} {\bibinfo {author} {\bibfnamefont {T.~M.}\ \bibnamefont
  {Lambley}, \bibfnamefont {H.~Schutzius}}\ and\ \bibinfo {author}
  {\bibfnamefont {D.}~\bibnamefont {Poulikakos}},\ }\bibfield  {title}
  {\enquote {\bibinfo {title} {Superhydrophobic surfaces for extreme
  environmental conditions},}\ }\href {\doibase 10.1073/pnas.2008775117}
  {\bibfield  {journal} {\bibinfo  {journal} {Proceedings of the National
  Academy of Sciences of the United States of America}\ }\textbf {\bibinfo
  {volume} {117}},\ \bibinfo {pages} {27188--27194} (\bibinfo {year}
  {2020})}\BibitemShut {NoStop}%
\bibitem [{\citenamefont {Tsai}\ \emph {et~al.}(2011)\citenamefont {Tsai},
  \citenamefont {Hendrix}, \citenamefont {Dijkstra}, \citenamefont {Shui},\
  and\ \citenamefont {Lohse}}]{Tsai2011MicroscopicStructure}%
  \BibitemOpen
  \bibfield  {author} {\bibinfo {author} {\bibfnamefont {P.~C.}\ \bibnamefont
  {Tsai}}, \bibinfo {author} {\bibfnamefont {M.~H.~W.}\ \bibnamefont
  {Hendrix}}, \bibinfo {author} {\bibfnamefont {R.~R.~M.}\ \bibnamefont
  {Dijkstra}}, \bibinfo {author} {\bibfnamefont {L.~L.}\ \bibnamefont {Shui}},
  \ and\ \bibinfo {author} {\bibfnamefont {D.}~\bibnamefont {Lohse}},\
  }\bibfield  {title} {\enquote {\bibinfo {title} {Microscopic structure
  influencing macroscopic splash at high {W}eber number},}\ }\href {\doibase
  10.1039/c1sm05801k} {\bibfield  {journal} {\bibinfo  {journal} {Soft Matter}\
  }\textbf {\bibinfo {volume} {7}},\ \bibinfo {pages} {11325--11333} (\bibinfo
  {year} {2011})}\BibitemShut {NoStop}%
\bibitem [{\citenamefont {Tsai}\ \emph {et~al.}(2010)\citenamefont {Tsai},
  \citenamefont {van~der Veen}, \citenamefont {van~de Raa},\ and\ \citenamefont
  {Lohse}}]{Tsai2010MicropatternAir}%
  \BibitemOpen
  \bibfield  {author} {\bibinfo {author} {\bibfnamefont {P.~C.}\ \bibnamefont
  {Tsai}}, \bibinfo {author} {\bibfnamefont {R.~C.~A.}\ \bibnamefont {van~der
  Veen}}, \bibinfo {author} {\bibfnamefont {M.}~\bibnamefont {van~de Raa}}, \
  and\ \bibinfo {author} {\bibfnamefont {D.}~\bibnamefont {Lohse}},\ }\bibfield
   {title} {\enquote {\bibinfo {title} {How micropatterns and air pressure
  affect splashing on surfaces},}\ }\href {\doibase 10.1021/la102330e}
  {\bibfield  {journal} {\bibinfo  {journal} {Langmuir}\ }\textbf {\bibinfo
  {volume} {26}},\ \bibinfo {pages} {16090--16095} (\bibinfo {year}
  {2010})}\BibitemShut {NoStop}%
\bibitem [{\citenamefont {Panao}\ and\ \citenamefont
  {Moreira}(2005)}]{Panao2005SprayImpingement}%
  \BibitemOpen
  \bibfield  {author} {\bibinfo {author} {\bibfnamefont {M.~R.~O.}\
  \bibnamefont {Panao}}\ and\ \bibinfo {author} {\bibfnamefont {A.~L.~N.}\
  \bibnamefont {Moreira}},\ }\bibfield  {title} {\enquote {\bibinfo {title}
  {Flow characteristics of spray impingement in {PFI} injection systems},}\
  }\href {\doibase 10.1007/s00348-005-0996-2} {\bibfield  {journal} {\bibinfo
  {journal} {Experiments in Fluids}\ }\textbf {\bibinfo {volume} {39}},\
  \bibinfo {pages} {364--374} (\bibinfo {year} {2005})}\BibitemShut {NoStop}%
\bibitem [{\citenamefont {Crockett}\ and\ \citenamefont
  {Horowitz}(2010)}]{Crockett2010ErosionNuclear}%
  \BibitemOpen
  \bibfield  {author} {\bibinfo {author} {\bibfnamefont {H.~M.}\ \bibnamefont
  {Crockett}}\ and\ \bibinfo {author} {\bibfnamefont {J.~S.}\ \bibnamefont
  {Horowitz}},\ }\bibfield  {title} {\enquote {\bibinfo {title} {Erosion in
  nuclear piping systems},}\ }\href {\doibase 10.1115/1.4000509} {\bibfield
  {journal} {\bibinfo  {journal} {Journal of Pressure Vessel
  Technology-Transactions of the Asme}\ }\textbf {\bibinfo {volume} {132}},\
  \bibinfo {pages} {024501} (\bibinfo {year} {2010})}\BibitemShut {NoStop}%
\bibitem [{\citenamefont {Sanchez-Caldera}, \citenamefont {Griffith},\ and\
  \citenamefont {Rabinowicz}(1988)}]{Sanchez1988CorrosionErosion}%
  \BibitemOpen
  \bibfield  {author} {\bibinfo {author} {\bibfnamefont {L.~E.}\ \bibnamefont
  {Sanchez-Caldera}}, \bibinfo {author} {\bibfnamefont {P.}~\bibnamefont
  {Griffith}}, \ and\ \bibinfo {author} {\bibfnamefont {E.}~\bibnamefont
  {Rabinowicz}},\ }\bibfield  {title} {\enquote {\bibinfo {title} {The
  mechanism of corrosion erosion in stream extraction lines of power
  stations},}\ }\href {\doibase 10.1115/1.3240099} {\bibfield  {journal}
  {\bibinfo  {journal} {Journal of Engineering for Gas Turbines and Power}\
  }\textbf {\bibinfo {volume} {110}},\ \bibinfo {pages} {180--184} (\bibinfo
  {year} {1988})}\BibitemShut {NoStop}%
\bibitem [{\citenamefont {Bischofberger}\ \emph {et~al.}(2016)\citenamefont
  {Bischofberger}, \citenamefont {Ray}, \citenamefont {Morris}, \citenamefont
  {Lee},\ and\ \citenamefont {Nagel}}]{Bischofberger2016AirflowGenerated}%
  \BibitemOpen
  \bibfield  {author} {\bibinfo {author} {\bibfnamefont {I.}~\bibnamefont
  {Bischofberger}}, \bibinfo {author} {\bibfnamefont {B.}~\bibnamefont {Ray}},
  \bibinfo {author} {\bibfnamefont {J.~F.}\ \bibnamefont {Morris}}, \bibinfo
  {author} {\bibfnamefont {T.}~\bibnamefont {Lee}}, \ and\ \bibinfo {author}
  {\bibfnamefont {S.~R.}\ \bibnamefont {Nagel}},\ }\bibfield  {title} {\enquote
  {\bibinfo {title} {Airflows generated by an impacting drop},}\ }\href
  {\doibase 10.1039/c5sm02702k} {\bibfield  {journal} {\bibinfo  {journal}
  {Soft Matter}\ }\textbf {\bibinfo {volume} {12}},\ \bibinfo {pages}
  {3013--3020} (\bibinfo {year} {2016})}\BibitemShut {NoStop}%
\bibitem [{\citenamefont {Shen}\ \emph {et~al.}(2016)\citenamefont {Shen},
  \citenamefont {Zou}, \citenamefont {Liu}, \citenamefont {Duley},\ and\
  \citenamefont {Zhou}}]{Shen2016MoltenDroplet}%
  \BibitemOpen
  \bibfield  {author} {\bibinfo {author} {\bibfnamefont {D.~Z.}\ \bibnamefont
  {Shen}}, \bibinfo {author} {\bibfnamefont {G.~S.}\ \bibnamefont {Zou}},
  \bibinfo {author} {\bibfnamefont {L.}~\bibnamefont {Liu}}, \bibinfo {author}
  {\bibfnamefont {W.~W.}\ \bibnamefont {Duley}}, \ and\ \bibinfo {author}
  {\bibfnamefont {Y.~N.}\ \bibnamefont {Zhou}},\ }\bibfield  {title} {\enquote
  {\bibinfo {title} {Investigation of splashing phenomena during the impact of
  molten sub-micron gold droplets on solid surfaces},}\ }\href {\doibase
  10.1039/c5sm00997a} {\bibfield  {journal} {\bibinfo  {journal} {Soft Matter}\
  }\textbf {\bibinfo {volume} {12}},\ \bibinfo {pages} {295--301} (\bibinfo
  {year} {2016})}\BibitemShut {NoStop}%
\bibitem [{\citenamefont {Song}\ \emph {et~al.}(2017)\citenamefont {Song},
  \citenamefont {Ju}, \citenamefont {Luo}, \citenamefont {Han}, \citenamefont
  {Dong}, \citenamefont {Wang}, \citenamefont {Gu}, \citenamefont {Zhang},
  \citenamefont {Hao},\ and\ \citenamefont
  {Jiang}}]{Song2017VesicleSurfactant}%
  \BibitemOpen
  \bibfield  {author} {\bibinfo {author} {\bibfnamefont {M.~R.}\ \bibnamefont
  {Song}}, \bibinfo {author} {\bibfnamefont {J.}~\bibnamefont {Ju}}, \bibinfo
  {author} {\bibfnamefont {S.~Q.}\ \bibnamefont {Luo}}, \bibinfo {author}
  {\bibfnamefont {Y.~C.}\ \bibnamefont {Han}}, \bibinfo {author} {\bibfnamefont
  {Z.~C.}\ \bibnamefont {Dong}}, \bibinfo {author} {\bibfnamefont {Y.~L.}\
  \bibnamefont {Wang}}, \bibinfo {author} {\bibfnamefont {Z.}~\bibnamefont
  {Gu}}, \bibinfo {author} {\bibfnamefont {L.~J.}\ \bibnamefont {Zhang}},
  \bibinfo {author} {\bibfnamefont {R.~R.}\ \bibnamefont {Hao}}, \ and\
  \bibinfo {author} {\bibfnamefont {L.}~\bibnamefont {Jiang}},\ }\bibfield
  {title} {\enquote {\bibinfo {title} {Controlling liquid splash on
  superhydrophobic surfaces by a vesicle surfactant},}\ }\href {\doibase
  10.1126/sciadv.1602188} {\bibfield  {journal} {\bibinfo  {journal} {Science
  Advances}\ }\textbf {\bibinfo {volume} {3}},\ \bibinfo {pages} {1602188}
  (\bibinfo {year} {2017})}\BibitemShut {NoStop}%
\bibitem [{\citenamefont {Visser}\ \emph {et~al.}(2015)\citenamefont {Visser},
  \citenamefont {Frommhold}, \citenamefont {Wildeman}, \citenamefont {Mettin},
  \citenamefont {Lohse},\ and\ \citenamefont
  {Sun}}]{Visser2015NumericalSimulation}%
  \BibitemOpen
  \bibfield  {author} {\bibinfo {author} {\bibfnamefont {C.~W.}\ \bibnamefont
  {Visser}}, \bibinfo {author} {\bibfnamefont {P.~E.}\ \bibnamefont
  {Frommhold}}, \bibinfo {author} {\bibfnamefont {S.}~\bibnamefont {Wildeman}},
  \bibinfo {author} {\bibfnamefont {R.}~\bibnamefont {Mettin}}, \bibinfo
  {author} {\bibfnamefont {D.}~\bibnamefont {Lohse}}, \ and\ \bibinfo {author}
  {\bibfnamefont {C.}~\bibnamefont {Sun}},\ }\bibfield  {title} {\enquote
  {\bibinfo {title} {Dynamics of high-speed micro-drop impact: numerical
  simulations and experiments at frame-to-frame times below 100 ns},}\ }\href
  {\doibase 10.1039/c4sm02474e} {\bibfield  {journal} {\bibinfo  {journal}
  {Soft Matter}\ }\textbf {\bibinfo {volume} {11}},\ \bibinfo {pages}
  {1708--1722} (\bibinfo {year} {2015})}\BibitemShut {NoStop}%
\bibitem [{\citenamefont
  {Worthington}(1876)}]{Worthington1876SymmetricSplashing}%
  \BibitemOpen
  \bibfield  {author} {\bibinfo {author} {\bibfnamefont {A.~M.}\ \bibnamefont
  {Worthington}},\ }\bibfield  {title} {\enquote {\bibinfo {title} {On the form
  assumed by drops of liquids falling vertically on a horizontal plate},}\
  }\href {\doibase 10.1098/rspl.1876.0048} {\bibfield  {journal} {\bibinfo
  {journal} {Proceedings of the Royal Society A-mathematical Physical and
  Engineering Sciences}\ }\textbf {\bibinfo {volume} {25}},\ \bibinfo {pages}
  {261--272} (\bibinfo {year} {1876})}\BibitemShut {NoStop}%
\bibitem [{\citenamefont {Mundo}, \citenamefont {Sommerfeld},\ and\
  \citenamefont {Tropea}(1995)}]{Mundo1995WallCollisions}%
  \BibitemOpen
  \bibfield  {author} {\bibinfo {author} {\bibfnamefont {C.}~\bibnamefont
  {Mundo}}, \bibinfo {author} {\bibfnamefont {M.}~\bibnamefont {Sommerfeld}}, \
  and\ \bibinfo {author} {\bibfnamefont {C.}~\bibnamefont {Tropea}},\
  }\bibfield  {title} {\enquote {\bibinfo {title} {Droplet-wall collisions:
  experimental studies of the deformation and breakup process},}\ }\href
  {\doibase 10.1016/0301-9322(94)00069-V} {\bibfield  {journal} {\bibinfo
  {journal} {International Journal of Multiphase Flow}\ }\textbf {\bibinfo
  {volume} {21}},\ \bibinfo {pages} {151--173} (\bibinfo {year}
  {1995})}\BibitemShut {NoStop}%
\bibitem [{\citenamefont {Range}\ and\ \citenamefont
  {Feuillebois}(1998)}]{Range1998SurfaceRoughness}%
  \BibitemOpen
  \bibfield  {author} {\bibinfo {author} {\bibfnamefont {K.}~\bibnamefont
  {Range}}\ and\ \bibinfo {author} {\bibfnamefont {F.}~\bibnamefont
  {Feuillebois}},\ }\bibfield  {title} {\enquote {\bibinfo {title} {Influence
  of surface roughness on liquid drop impact},}\ }\href {\doibase
  10.1006/jcis.1998.5518} {\bibfield  {journal} {\bibinfo  {journal} {Journal
  of Colloid and Interface Science}\ }\textbf {\bibinfo {volume} {203}},\
  \bibinfo {pages} {16--30} (\bibinfo {year} {1998})}\BibitemShut {NoStop}%
\bibitem [{\citenamefont {Wang}\ and\ \citenamefont
  {Bourouiba}(2018)}]{Wang2018SheetFragmentation}%
  \BibitemOpen
  \bibfield  {author} {\bibinfo {author} {\bibfnamefont {Y.}~\bibnamefont
  {Wang}}\ and\ \bibinfo {author} {\bibfnamefont {L.}~\bibnamefont
  {Bourouiba}},\ }\bibfield  {title} {\enquote {\bibinfo {title} {Unsteady
  sheet fragmentation: droplet sizes and speeds},}\ }\href {\doibase
  10.1017/jfm.2018.359} {\bibfield  {journal} {\bibinfo  {journal} {Journal of
  Fluid Mechanics}\ }\textbf {\bibinfo {volume} {848}},\ \bibinfo {pages}
  {946--967} (\bibinfo {year} {2018})}\BibitemShut {NoStop}%
\bibitem [{\citenamefont {Zhang}\ \emph {et~al.}(2010)\citenamefont {Zhang},
  \citenamefont {Brunet}, \citenamefont {Eggers},\ and\ \citenamefont
  {Deegan}}]{Zhang2010WavelengthSelection}%
  \BibitemOpen
  \bibfield  {author} {\bibinfo {author} {\bibfnamefont {L.~V.}\ \bibnamefont
  {Zhang}}, \bibinfo {author} {\bibfnamefont {P.}~\bibnamefont {Brunet}},
  \bibinfo {author} {\bibfnamefont {J.}~\bibnamefont {Eggers}}, \ and\ \bibinfo
  {author} {\bibfnamefont {R.~D.}\ \bibnamefont {Deegan}},\ }\bibfield  {title}
  {\enquote {\bibinfo {title} {Wavelength selection in the crown splash},}\
  }\href {\doibase 10.1063/1.3526743} {\bibfield  {journal} {\bibinfo
  {journal} {Physics of Fluids}\ }\textbf {\bibinfo {volume} {22}},\ \bibinfo
  {pages} {122105} (\bibinfo {year} {2010})}\BibitemShut {NoStop}%
\bibitem [{\citenamefont {Wang}\ \emph {et~al.}(2018)\citenamefont {Wang},
  \citenamefont {Dandekar}, \citenamefont {Bustos}, \citenamefont {Poulain},\
  and\ \citenamefont {Bourouiba}}]{Wang2018UnsteadyFragmentation}%
  \BibitemOpen
  \bibfield  {author} {\bibinfo {author} {\bibfnamefont {Y.}~\bibnamefont
  {Wang}}, \bibinfo {author} {\bibfnamefont {R.}~\bibnamefont {Dandekar}},
  \bibinfo {author} {\bibfnamefont {N.}~\bibnamefont {Bustos}}, \bibinfo
  {author} {\bibfnamefont {S.}~\bibnamefont {Poulain}}, \ and\ \bibinfo
  {author} {\bibfnamefont {L.}~\bibnamefont {Bourouiba}},\ }\bibfield  {title}
  {\enquote {\bibinfo {title} {Universal rim thickness in unsteady sheet
  fragmentation},}\ }\href {\doibase 10.1103/PhysRevLett.120.204503} {\bibfield
   {journal} {\bibinfo  {journal} {Physical Review Letters}\ }\textbf {\bibinfo
  {volume} {120}},\ \bibinfo {pages} {204503} (\bibinfo {year}
  {2018})}\BibitemShut {NoStop}%
\bibitem [{\citenamefont {Wang}\ and\ \citenamefont
  {Bourouiba}(2021)}]{Wang2021UnsteadyFragmentation}%
  \BibitemOpen
  \bibfield  {author} {\bibinfo {author} {\bibfnamefont {Y.}~\bibnamefont
  {Wang}}\ and\ \bibinfo {author} {\bibfnamefont {L.}~\bibnamefont
  {Bourouiba}},\ }\bibfield  {title} {\enquote {\bibinfo {title} {Growth and
  breakup of ligaments in unsteady fragmentation},}\ }\href {\doibase
  10.1017/jfm.2020.698} {\bibfield  {journal} {\bibinfo  {journal} {Journal of
  Fluid Mechanics}\ }\textbf {\bibinfo {volume} {910}},\ \bibinfo {pages} {698}
  (\bibinfo {year} {2021})}\BibitemShut {NoStop}%
\bibitem [{\citenamefont {Roisman}, \citenamefont {Horvat},\ and\ \citenamefont
  {Tropea}(2006)}]{Roisman2006FingeringSplashing}%
  \BibitemOpen
  \bibfield  {author} {\bibinfo {author} {\bibfnamefont {I.~V.}\ \bibnamefont
  {Roisman}}, \bibinfo {author} {\bibfnamefont {K.}~\bibnamefont {Horvat}}, \
  and\ \bibinfo {author} {\bibfnamefont {C.}~\bibnamefont {Tropea}},\
  }\bibfield  {title} {\enquote {\bibinfo {title} {Spray impact: rim transverse
  instability initiating fingering and splash, and description of a secondary
  spray},}\ }\href {\doibase 10.1063/1.2364187} {\bibfield  {journal} {\bibinfo
   {journal} {Physics of Fluids}\ }\textbf {\bibinfo {volume} {18}},\ \bibinfo
  {pages} {102104} (\bibinfo {year} {2006})}\BibitemShut {NoStop}%
\bibitem [{\citenamefont {Villermaux}\ and\ \citenamefont
  {Bossa}(2011)}]{Villermaux2011DropFragmentation}%
  \BibitemOpen
  \bibfield  {author} {\bibinfo {author} {\bibfnamefont {E.}~\bibnamefont
  {Villermaux}}\ and\ \bibinfo {author} {\bibfnamefont {B.}~\bibnamefont
  {Bossa}},\ }\bibfield  {title} {\enquote {\bibinfo {title} {Drop
  fragmentation on impact},}\ }\href {\doibase 10.1017/S002211201000474x}
  {\bibfield  {journal} {\bibinfo  {journal} {Journal of Fluid Mechanics}\
  }\textbf {\bibinfo {volume} {668}},\ \bibinfo {pages} {412--435} (\bibinfo
  {year} {2011})}\BibitemShut {NoStop}%
\bibitem [{\citenamefont {Thoroddsen}\ \emph {et~al.}(2011)\citenamefont
  {Thoroddsen}, \citenamefont {Thoraval}, \citenamefont {Takehara},\ and\
  \citenamefont {Etoh}}]{Thoroddsen2011SlingshotMechanism}%
  \BibitemOpen
  \bibfield  {author} {\bibinfo {author} {\bibfnamefont {S.~T.}\ \bibnamefont
  {Thoroddsen}}, \bibinfo {author} {\bibfnamefont {M.~J.}\ \bibnamefont
  {Thoraval}}, \bibinfo {author} {\bibfnamefont {K.}~\bibnamefont {Takehara}},
  \ and\ \bibinfo {author} {\bibfnamefont {T.~G.}\ \bibnamefont {Etoh}},\
  }\bibfield  {title} {\enquote {\bibinfo {title} {Droplet splashing by a
  slingshot mechanism},}\ }\href {\doibase 10.1103/PhysRevLett.106.034501}
  {\bibfield  {journal} {\bibinfo  {journal} {Physical Review Letters}\
  }\textbf {\bibinfo {volume} {106}},\ \bibinfo {pages} {034501} (\bibinfo
  {year} {2011})}\BibitemShut {NoStop}%
\bibitem [{\citenamefont {Driscoll}, \citenamefont {Stevens},\ and\
  \citenamefont {Nagel}(2010)}]{Driscoll2010FilmFormation}%
  \BibitemOpen
  \bibfield  {author} {\bibinfo {author} {\bibfnamefont {M.~M.}\ \bibnamefont
  {Driscoll}}, \bibinfo {author} {\bibfnamefont {C.~S.}\ \bibnamefont
  {Stevens}}, \ and\ \bibinfo {author} {\bibfnamefont {S.~R.}\ \bibnamefont
  {Nagel}},\ }\bibfield  {title} {\enquote {\bibinfo {title} {Thin film
  formation during splashing of viscous liquids},}\ }\href {\doibase
  10.1103/PhysRevE.82.036302} {\bibfield  {journal} {\bibinfo  {journal}
  {Physical Review E}\ }\textbf {\bibinfo {volume} {82}},\ \bibinfo {pages}
  {036302} (\bibinfo {year} {2010})}\BibitemShut {NoStop}%
\bibitem [{\citenamefont {Thoroddsen}, \citenamefont {Takehara},\ and\
  \citenamefont {Etoh}(2010)}]{Thoroddsen2010BubbleEntrapment}%
  \BibitemOpen
  \bibfield  {author} {\bibinfo {author} {\bibfnamefont {S.}~\bibnamefont
  {Thoroddsen}}, \bibinfo {author} {\bibfnamefont {K.}~\bibnamefont
  {Takehara}}, \ and\ \bibinfo {author} {\bibfnamefont {T.}~\bibnamefont
  {Etoh}},\ }\bibfield  {title} {\enquote {\bibinfo {title} {Bubble entrapment
  through topological change},}\ }\href {\doibase 10.1063/1.3407654} {\bibfield
   {journal} {\bibinfo  {journal} {Physics of Fluids}\ }\textbf {\bibinfo
  {volume} {22}},\ \bibinfo {pages} {051701} (\bibinfo {year}
  {2010})}\BibitemShut {NoStop}%
\bibitem [{\citenamefont {Thoroddsen}, \citenamefont {Takehara},\ and\
  \citenamefont {Etoh}(2012)}]{Thoroddsen2012Splashing}%
  \BibitemOpen
  \bibfield  {author} {\bibinfo {author} {\bibfnamefont {S.~T.}\ \bibnamefont
  {Thoroddsen}}, \bibinfo {author} {\bibfnamefont {K.}~\bibnamefont
  {Takehara}}, \ and\ \bibinfo {author} {\bibfnamefont {T.~G.}\ \bibnamefont
  {Etoh}},\ }\bibfield  {title} {\enquote {\bibinfo {title} {Micro-splashing by
  drop impacts},}\ }\href {\doibase 10.1017/jfm.2012.281} {\bibfield  {journal}
  {\bibinfo  {journal} {Journal of Fluid Mechanics}\ }\textbf {\bibinfo
  {volume} {706}},\ \bibinfo {pages} {560--570} (\bibinfo {year}
  {2012})}\BibitemShut {NoStop}%
\bibitem [{\citenamefont {Xu}(2010)}]{Xu2010InstabilityImpacting}%
  \BibitemOpen
  \bibfield  {author} {\bibinfo {author} {\bibfnamefont {L.}~\bibnamefont
  {Xu}},\ }\bibfield  {title} {\enquote {\bibinfo {title} {Instability
  development of a viscous liquid drop impacting a smooth substrate},}\ }\href
  {\doibase 10.1103/PhysRevE.82.025303} {\bibfield  {journal} {\bibinfo
  {journal} {Physical Review E}\ }\textbf {\bibinfo {volume} {82}},\ \bibinfo
  {pages} {025303} (\bibinfo {year} {2010})}\BibitemShut {NoStop}%
\bibitem [{\citenamefont {Stevens}, \citenamefont {Latka},\ and\ \citenamefont
  {Nagel}(2014)}]{Stevens2014SplashingViscosity}%
  \BibitemOpen
  \bibfield  {author} {\bibinfo {author} {\bibfnamefont {C.~S.}\ \bibnamefont
  {Stevens}}, \bibinfo {author} {\bibfnamefont {A.}~\bibnamefont {Latka}}, \
  and\ \bibinfo {author} {\bibfnamefont {S.~R.}\ \bibnamefont {Nagel}},\
  }\bibfield  {title} {\enquote {\bibinfo {title} {Comparison of splashing in
  high- and low-viscosity liquids},}\ }\href {\doibase
  10.1103/PhysRevE.89.063006} {\bibfield  {journal} {\bibinfo  {journal}
  {Physical Review E}\ }\textbf {\bibinfo {volume} {89}},\ \bibinfo {pages}
  {063006} (\bibinfo {year} {2014})}\BibitemShut {NoStop}%
\bibitem [{\citenamefont {Zhang}\ \emph {et~al.}(2021)\citenamefont {Zhang},
  \citenamefont {Zhang}, \citenamefont {Yi}, \citenamefont {He}, \citenamefont
  {Niu},\ and\ \citenamefont {Hao}}]{Zhang2021ReversedViscositSplash}%
  \BibitemOpen
  \bibfield  {author} {\bibinfo {author} {\bibfnamefont {H.~X.}\ \bibnamefont
  {Zhang}}, \bibinfo {author} {\bibfnamefont {X.~W.}\ \bibnamefont {Zhang}},
  \bibinfo {author} {\bibfnamefont {X.}~\bibnamefont {Yi}}, \bibinfo {author}
  {\bibfnamefont {F.}~\bibnamefont {He}}, \bibinfo {author} {\bibfnamefont
  {F.~L.}\ \bibnamefont {Niu}}, \ and\ \bibinfo {author} {\bibfnamefont
  {P.~F.}\ \bibnamefont {Hao}},\ }\bibfield  {title} {\enquote {\bibinfo
  {title} {Reversed role of liquid viscosity on drop splash},}\ }\href
  {\doibase 10.1063/5.0048569} {\bibfield  {journal} {\bibinfo  {journal}
  {Physics of Fluids}\ }\textbf {\bibinfo {volume} {33}},\ \bibinfo {pages}
  {052103} (\bibinfo {year} {2021})}\BibitemShut {NoStop}%
\bibitem [{\citenamefont {Liu}\ \emph {et~al.}(2010)\citenamefont {Liu},
  \citenamefont {Vu}, \citenamefont {Yoon}, \citenamefont {Jepsen},\ and\
  \citenamefont {Aguilar}}]{Liu2010SplashingInclined}%
  \BibitemOpen
  \bibfield  {author} {\bibinfo {author} {\bibfnamefont {J.}~\bibnamefont
  {Liu}}, \bibinfo {author} {\bibfnamefont {H.}~\bibnamefont {Vu}}, \bibinfo
  {author} {\bibfnamefont {S.~S.}\ \bibnamefont {Yoon}}, \bibinfo {author}
  {\bibfnamefont {R.}~\bibnamefont {Jepsen}}, \ and\ \bibinfo {author}
  {\bibfnamefont {G.}~\bibnamefont {Aguilar}},\ }\bibfield  {title} {\enquote
  {\bibinfo {title} {Splashing phenomena during liquid droplet impact},}\
  }\href {\doibase 10.1615/AtomizSpr.v20.i4.30} {\bibfield  {journal} {\bibinfo
   {journal} {Atomization and Sprays}\ }\textbf {\bibinfo {volume} {20}},\
  \bibinfo {pages} {297--310} (\bibinfo {year} {2010})}\BibitemShut {NoStop}%
\bibitem [{\citenamefont {Agbaglah}, \citenamefont {Josserand},\ and\
  \citenamefont {Zaleski}(2013)}]{Agbaglah2013InstabilityRim}%
  \BibitemOpen
  \bibfield  {author} {\bibinfo {author} {\bibfnamefont {G.}~\bibnamefont
  {Agbaglah}}, \bibinfo {author} {\bibfnamefont {C.}~\bibnamefont {Josserand}},
  \ and\ \bibinfo {author} {\bibfnamefont {S.}~\bibnamefont {Zaleski}},\
  }\bibfield  {title} {\enquote {\bibinfo {title} {Longitudinal instability of
  a liquid rim},}\ }\href {\doibase 10.1063/1.4789971} {\bibfield  {journal}
  {\bibinfo  {journal} {Physics of Fluids}\ }\textbf {\bibinfo {volume} {25}},\
  \bibinfo {pages} {022103} (\bibinfo {year} {2013})}\BibitemShut {NoStop}%
\bibitem [{\citenamefont {Rayleigh}(1878)}]{Rayleigh1878OnInstabilityJets}%
  \BibitemOpen
  \bibfield  {author} {\bibinfo {author} {\bibfnamefont {L.}~\bibnamefont
  {Rayleigh}},\ }\bibfield  {title} {\enquote {\bibinfo {title} {On the
  instability of jets},}\ }\href {\doibase 10.1112/plms/s1-10.1.4} {\bibfield
  {journal} {\bibinfo  {journal} {Proceedings of the London Mathematical
  Society}\ }\textbf {\bibinfo {volume} {s1-10}},\ \bibinfo {pages} {4--13}
  (\bibinfo {year} {1878})}\BibitemShut {NoStop}%
\bibitem [{\citenamefont {Taylor}(1950)}]{Taylor1950InstabilityLiquidSurfaces}%
  \BibitemOpen
  \bibfield  {author} {\bibinfo {author} {\bibfnamefont {G.}~\bibnamefont
  {Taylor}},\ }\bibfield  {title} {\enquote {\bibinfo {title} {The instability
  of liquid surfaces when accelerated in a direction perpendicular to their
  planes. i},}\ }\href {\doibase doi:10.1098/rspa.1950.0052} {\bibfield
  {journal} {\bibinfo  {journal} {Proceedings of the London Mathematical
  Society}\ }\textbf {\bibinfo {volume} {201}},\ \bibinfo {pages} {192--196}
  (\bibinfo {year} {1950})}\BibitemShut {NoStop}%
\bibitem [{\citenamefont {Yarin}\ and\ \citenamefont
  {Weiss}(1995)}]{Yarin1995CapillaryWaves}%
  \BibitemOpen
  \bibfield  {author} {\bibinfo {author} {\bibfnamefont {A.}~\bibnamefont
  {Yarin}}\ and\ \bibinfo {author} {\bibfnamefont {D.}~\bibnamefont {Weiss}},\
  }\bibfield  {title} {\enquote {\bibinfo {title} {Impact of drops on solid
  surfaces: self-similar capillary waves, and splashing as a new type of
  kinematic discontinuity},}\ }\href {\doibase 10.1017/S0022112095002266}
  {\bibfield  {journal} {\bibinfo  {journal} {Journal of Fluid Mechanics}\
  }\textbf {\bibinfo {volume} {283}},\ \bibinfo {pages} {141--152} (\bibinfo
  {year} {1995})}\BibitemShut {NoStop}%
\bibitem [{\citenamefont {Riboux}\ and\ \citenamefont
  {Gordillo}(2014)}]{Riboux2014CriticalSpeed}%
  \BibitemOpen
  \bibfield  {author} {\bibinfo {author} {\bibfnamefont {G.}~\bibnamefont
  {Riboux}}\ and\ \bibinfo {author} {\bibfnamefont {J.~M.}\ \bibnamefont
  {Gordillo}},\ }\bibfield  {title} {\enquote {\bibinfo {title} {Experiments of
  drops impacting a smooth solid surface: a model of the critical impact speed
  for drop splashing},}\ }\href {\doibase 10.1103/PhysRevLett.113.024507}
  {\bibfield  {journal} {\bibinfo  {journal} {Physical Review Letters}\
  }\textbf {\bibinfo {volume} {113}},\ \bibinfo {pages} {024507} (\bibinfo
  {year} {2014})}\BibitemShut {NoStop}%
\bibitem [{\citenamefont {Ashida}\ \emph {et~al.}(2020)\citenamefont {Ashida},
  \citenamefont {Watanabe}, \citenamefont {Kobayashi}, \citenamefont {Fujii},\
  and\ \citenamefont {Sanada}}]{Ashida2020PromptSplashing}%
  \BibitemOpen
  \bibfield  {author} {\bibinfo {author} {\bibfnamefont {T.}~\bibnamefont
  {Ashida}}, \bibinfo {author} {\bibfnamefont {M.}~\bibnamefont {Watanabe}},
  \bibinfo {author} {\bibfnamefont {K.}~\bibnamefont {Kobayashi}}, \bibinfo
  {author} {\bibfnamefont {H.}~\bibnamefont {Fujii}}, \ and\ \bibinfo {author}
  {\bibfnamefont {T.}~\bibnamefont {Sanada}},\ }\bibfield  {title} {\enquote
  {\bibinfo {title} {Hidden prompt splashing by corona splashing at drop impact
  on a smooth dry surface},}\ }\href {\doibase 10.1103/PhysRevFluids.5.011601}
  {\bibfield  {journal} {\bibinfo  {journal} {Physical Review Fluids}\ }\textbf
  {\bibinfo {volume} {5}},\ \bibinfo {pages} {011601} (\bibinfo {year}
  {2020})}\BibitemShut {NoStop}%
\bibitem [{\citenamefont {Gordillo}\ and\ \citenamefont
  {Riboux}(2019)}]{Gordillo2019AerodynamicSplashing}%
  \BibitemOpen
  \bibfield  {author} {\bibinfo {author} {\bibfnamefont {J.~M.}\ \bibnamefont
  {Gordillo}}\ and\ \bibinfo {author} {\bibfnamefont {G.}~\bibnamefont
  {Riboux}},\ }\bibfield  {title} {\enquote {\bibinfo {title} {A note on the
  aerodynamic splashing of droplets},}\ }\href {\doibase 10.1017/jfm.2019.396}
  {\bibfield  {journal} {\bibinfo  {journal} {Journal of Fluid Mechanics}\
  }\textbf {\bibinfo {volume} {871}},\ \bibinfo {pages} {R3} (\bibinfo {year}
  {2019})}\BibitemShut {NoStop}%
\bibitem [{\citenamefont {Hao}\ \emph {et~al.}(2019)\citenamefont {Hao},
  \citenamefont {Lu}, \citenamefont {Lee}, \citenamefont {Wu}, \citenamefont
  {Hu},\ and\ \citenamefont {Floryan}}]{Hao2019InclinedSurface}%
  \BibitemOpen
  \bibfield  {author} {\bibinfo {author} {\bibfnamefont {J.~G.}\ \bibnamefont
  {Hao}}, \bibinfo {author} {\bibfnamefont {J.}~\bibnamefont {Lu}}, \bibinfo
  {author} {\bibfnamefont {L.~N.}\ \bibnamefont {Lee}}, \bibinfo {author}
  {\bibfnamefont {Z.~H.}\ \bibnamefont {Wu}}, \bibinfo {author} {\bibfnamefont
  {G.~K.}\ \bibnamefont {Hu}}, \ and\ \bibinfo {author} {\bibfnamefont {J.~M.}\
  \bibnamefont {Floryan}},\ }\bibfield  {title} {\enquote {\bibinfo {title}
  {Droplet splashing on an inclined surface},}\ }\href {\doibase
  10.1103/PhysRevLett.122.054501} {\bibfield  {journal} {\bibinfo  {journal}
  {Physical Review Letters}\ }\textbf {\bibinfo {volume} {122}},\ \bibinfo
  {pages} {054501} (\bibinfo {year} {2019})}\BibitemShut {NoStop}%
\end{thebibliography}%
\end{document}